\documentclass[aps,preprint,nofootinbib,floatfix]{revtex4}
\usepackage{epsfig}
\usepackage{amsmath,amsfonts}
\usepackage{subfigure}

\def\Tr{\mathrm{Tr}}
\def\Str{\mathrm{Str}}
\def\chpt{\raise0.4ex\hbox{$\chi$}PT}

\def\CL{{\cal L}}

\def\CS{{\cal S}}

\def\CV{{\cal V}}

\def\Dslash{{\rm D}\!\!\!\!/\,}
\def\hat{\widehat}
\def\eff{{\rm eff}}
\def\aux{{\rm aux}}
\def\rep{{\rm rep}}
\def\min{{\rm min}}

\def\spose#1{\hbox to 0pt{#1\hss}}
\def\ltapprox{\mathrel{\spose{\lower 3pt\hbox{$\mathchar"218$}}
 \raise 2.0pt\hbox{$\mathchar"13C$}}}
\def\gtapprox{\mathrel{\spose{\lower 3pt\hbox{$\mathchar"218$}}
 \raise 2.0pt\hbox{$\mathchar"13E$}}}
\def\inapprox{\mathrel{\spose{\lower 3pt\hbox{$\mathchar"218$}}
 \raise 2.0pt\hbox{$\mathchar"232$}}}

\begin{document}

\title{Discretization errors in the spectrum of the
Hermitian Wilson-Dirac operator}

\author{Stephen~R.~Sharpe}
\email[]{sharpe@phys.washington.edu}
\affiliation{Physics Department, University of Washington,
Seattle, WA 98195-1560, USA}

\date{\today}

\begin{abstract}
I study the leading effects of discretization errors on the
low-energy part of the 
spectrum of the Hermitian Wilson-Dirac operator in infinite volume.
The method generalizes that used to study the spectrum of the Dirac
operator in the continuum, and uses
partially quenched chiral perturbation theory for Wilson fermions.
The leading-order corrections are proportional to $a^2$
($a$ being the lattice spacing).
At this order I find that the method works only for
one choice of sign of one of the three low-energy constants describing
discretization errors. If these constants have the relative magnitudes 
expected from
large $N_c$ arguments, then the method works if the theory has an Aoki phase
for $m\sim a^2$, but fails if there is a first-order transition.
In the former case, the dependence of the gap and the spectral
density on $m$ and $a^2$ are determined. In particular, the gap is found to
vanish more quickly as $m_\pi^2\to 0$ than in the continuum. This reduces
the region where simulations are safe from fluctuations in the gap.
\end{abstract}

\maketitle

\section{\label{sec:outline} Introduction}

Simulations of lattice QCD using Wilson fermions are now able to enter
into the chiral regime~\cite{Wilsonchiral}, in large part due to recent 
advances in algorithms~\cite{SAP,Hasenbusch,Urbach}. An obstacle to
further reduction is the possible presence of arbitrarily small eigenvalues
of the Wilson-Dirac operator, due to the breaking of chiral symmetry.
Such small eigenvalues can lead to algorithmic instabilities and slowing 
down.\footnote{%
For a summary of possible problems, see Ref.~\cite{DD}.}
This issue has been investigated numerically in Ref.~\cite{DD}, where 
it is found that
the average spectral gap is approximately proportional to the quark mass,
with a distribution whose width
is $\approx a/\sqrt{V}$, with $a$ the lattice spacing and
$V$ the four-volume of the lattice. 
This allows one to estimate the parameters for which simulations
are safe from instabilities.



These results are encouraging, but raise further questions.
In particular, what are the leading-order (LO) effects of
discretization on the spectrum which survive in infinite volume?
Do such effects impact the regime in which light quarks
can be safely simulated?
This paper aims to address these questions, at least in part,
by determining the leading effect of discretization errors 
on the infinite volume spectral density.

In fact, it is already known that the
spectral density in infinite volume is distorted by
discretization errors proportional to $a^2$.
Such errors lead to a non-trivial phase structure
in the regime where $m\sim a^2\Lambda_{\rm QCD}^3$~\cite{Creutz,ShSi},
in which there is competition between chiral symmetry breaking due to 
quark masses (generically denoted $m$) and the Wilson term.
In the scenario with an Aoki phase~\cite{Aoki}, it is found that
the gap vanishes throughout the Aoki phase, and thus
for a range of quark masses, rather than for a single value ($m=0$) as in
the continuum~\cite{ShSi}.
The results obtained here quantify this further, by showing how the gap
approaches zero as one approaches the Aoki-phase end points,
and determining the distortion of the spectral density
from its continuum form.

The above-mentioned competition between quark mass and
$a^2$ effects can lead to another scenario, in which there is no Aoki phase,
but instead a first-order phase transition 
at a non-zero pion mass~\cite{Creutz,ShSi}.
Pions lighter than this minimal mass are not accessible.
One does not know {\em a priori} which scenario holds for 
a particular choice of action.
Numerical simulations find, however, that the first-order scenario appears to apply
for simulations with (unimproved) Wilson fermions and the Wilson gauge action 
for $a\ltapprox 0.2\;$fm~\cite{firstorder}.
The resulting minimal pion mass is surprisingly large, roughly $500\;$MeV at
$a\approx 0.2\;$fm. In terms of the leading-order prediction,
$(m_\pi^{\rm min})^2= a^2 \Lambda^4$, this corresponds to a non-perturbative
scale $\Lambda \approx 700\;$MeV, a large but not unreasonable value.
The large size of this effect provides further motivation for studying the
discretization errors in the spectral density.\footnote{%
One might wonder if this large value of $\Lambda$ is consistent with the results
of Refs.~\cite{Wilsonchiral,DD}, which use Wilson fermions and the Wilson gauge 
action, and work at small pion masses and yet see no first-order transition.
In fact, it is consistent because the minimal pion masses
($\approx 250,\;200\ {\rm and}\ 150\;$MeV at 
$a=0.1,\; 0.08\ {\rm and}\ 0.06\;$fm, respectively)
lie below the smallest pion masses simulated.}
%

I use two methods to study the spectral density and gap.
Both rely on partially quenched (PQ) chiral perturbation theory (\chpt)
including the effects of discretization errors.
The first method (which I call the ``primary method'') makes
less use of partial quenching, and appears to be on more solid
theoretical ground.
It turns out, however, that it fails at leading-order 
(the order I work at here) for certain choices of
the parameters (``low-energy constants'' or LECs)
in the chiral Lagrangian. These correspond approximately to
the parameters for which there is a first-order transition. Thus the method 
works predominantly if the theory is in the scenario with an Aoki phase.
Due to this limitation,
it will probably not be applicable to simulations with
unimproved Wilson quarks and the Wilson gauge action. 
It may, however, be applicable
to simulations involving improved 
Wilson fermions and/or improved gauge actions,
where the nature of the phase diagram is not yet clear.

Because of the limitations of the primary method,
and as a check, I have repeated the calculation
 using a different method (which I call the ``alternate method'').
This method requires more extensive use of the properties of PQ
 chiral theories, properties I study using the replica trick.
I find that the successes of the first method are reproduced, 
but that, in general, its failures are not resolved,
although some avenues for further study are suggested.

The remainder of this paper is organized as follows. In the next section 
I give an overview of the 
methods I use to determine the spectral density.
In sec.~\ref{sec:chpt}, I review
PQ\chpt\ including discretization effects.
The derivation of the primary method is given in sec.~\ref{sec:chptcalc},
and the results of this method are presented in sec.~\ref{sec:results}.
Discussion and conclusions appear in sec.~\ref{sec:conc}.
 Appendix~\ref{app:key}
gives a derivation of a key result used in sec.~\ref{sec:chptcalc}.
Appendix~\ref{app:replica} describes the calculation using the
alternate method.

\section{\label{sec:method} Overview of Methods}

The low-energy part of the spectrum of the Wilson-Dirac operator is related to
long-distance physics, the classic example in the continuum being the Banks-Casher
relation between the density of zero eigenvalues and the quark 
condensate~\cite{BanksCasher}.
The basic idea of the method used here is to represent
 long-distance physics by the effective
chiral Lagrangian, into which the effects of discretization can be systematically
introduced~\cite{ShSi,RuSh,BRS}. The aim is then to convert results
from the chiral Lagrangian concerning
the long-distance behavior of correlators of bilinear operators
into a prediction for the spectral density of the lattice operator.

A tool for doing so was developed for the continuum theory
by Refs.~\cite{Osborn,Damgaard}.
One introduces a valence quark, the condensate for which has the
spectral decomposition:
\begin{equation}
\langle \bar q_V q_V\rangle(m_V) = - \int d\lambda 
\frac{\rho_{\Dslash}(\lambda)}{i\lambda+m_V} \,,
\label{eq:condV}
\end{equation}
where $\rho_{\Dslash}(\lambda)$ is the spectral density of the 
hermitian continuum operator $-i\Dslash$.
The key point is that this spectral density is independent of the valence
quark mass, since it does not enter the quark determinant. Thus the
relation can be inverted 
(unlike the corresponding expression for the sea-quark determinant):
\begin{equation}
{\rm Disc}[\langle\bar q_V q_V\rangle]\bigg|_{m_V=-i\lambda} 
= - 2 \pi \rho_{\Dslash}(\lambda) \,,
\label{eq:osborn}
\end{equation}
so that one needs to calculate the valence condensate 
for complex valence quark mass.
This can be done by analytically continuing the expressions obtained using
PQ\chpt~\cite{BGPQ} for real masses. 

The leading-order analysis is very simple~\cite{Osborn}. 
Assume that the sea-quark masses are small and positive, so the sea-quark
condensate $\langle\bar q_S q_S\rangle$ is negative.
Then, for real $m_V$, the valence condensate
has the same magnitude as that of the sea-quark, but with
a sign given by that of $m_V$.\footnote{%
This result is demonstrated in passing in Appendix~\ref{app:replica}
using the replica trick.}
Analytically continuing to complex $m_V$ gives
\begin{equation}
\langle\bar q_V q_V\rangle 
= \langle\bar q_S q_S\rangle \sqrt{m_V^2}/m_V \,.
\label{eq:LOcont}
\end{equation}
The discontinuity for imaginary $m_V$ gives, 
using eq.~(\ref{eq:osborn}), the Banks-Casher relation
\begin{equation}
\rho_{\Dslash}(\lambda)= -\frac{\langle\bar q_S q_S\rangle}{\pi} 
\left[1 + O(|\lambda|/\Lambda_{\rm QCD})\right]
\,.
\label{eq:rhofromBC}
\end{equation}
The NLO correction (the term linear in $|\lambda|$) was
also obtained in Ref.~\cite{Osborn}
(reproducing and extending earlier results of Ref.~\cite{SternSmilga}),
and the calculation was extended
to the $\epsilon$-regime in Ref.~\cite{Damgaard},
giving the microscopic spectral density.

My aim in this paper is to generalize this simple continuum LO
analysis to the lattice theory. This turns out to be more
involved than one might have anticipated.
Using the most direct generalization of eq.~(\ref{eq:osborn}) leads
to complications which I only partly resolve in Appendix~\ref{app:replica}. 
This approach I name the ``alternate method''. 
Instead, in the main text I follow a different, though related, 
approach, which I call the ``primary method''.
In the following, I first present the key equation
of the alternate method since this provides useful background
for the subsequent development of the primary method.

To generalize eq.~(\ref{eq:osborn}),
one cannot simply replace $\Dslash$ with the Wilson-Dirac
operator $D_W$, since the latter is not anti-hermitian,
and has complex eigenvalues. Instead, the spectrum of interest
is that of the hermitian Wilson-Dirac operator
\begin{equation}
Q_m=\gamma_5 (D_W+m_0) = Q_m^\dagger\,,
\end{equation}
with $m_0$ the bare quark mass.
To generalize eq.~(\ref{eq:condV}) I
note that to introduce the $\gamma_5$ needed to convert
the propagator $(D_W+m_0)^{-1}$ to the desired $Q_m^{-1}$,
one should consider the valence {\em pseudoscalar} condensate.
Furthermore, the extra $\gamma_5$ in $Q_m$ means that the analog
of $m_V$ in (\ref{eq:condV}) is a valence {\em twisted} mass. 
I find it convenient to
introduce a degenerate pair of valence quarks, for which one
can write a flavor non-singlet twisted
mass term as $i\mu \bar q_V \gamma_5 \tau_3 q_V$.
It is then straightforward to derive an analog of (\ref{eq:condV}), namely
\begin{equation}
\langle \bar q_V \gamma_5 \tau_3 q_V\rangle(\mu) = - \int d\lambda 
\frac{\left[\rho_{Q}(\lambda)+\rho_{Q}(-\lambda)\right]}{\lambda+ i \mu} \,,
\end{equation}
where $\rho_Q(\lambda)$ is the spectral density of $Q_m$.
As before, the use of a valence quark allows one to invert this relationship, 
yielding
\begin{equation}
{\rm Disc}[\langle\bar q_V \gamma_5 \tau_3 q_V\rangle]\bigg|_{\mu=i\lambda}  
= 2 i \pi \left[\rho_{Q}(\lambda)+\rho_{Q}(-\lambda)\right]
\,.
\label{eq:PQapproach}
\end{equation}
Note that the {\em untwisted} part of the bare valence quark mass is
to be kept at the same value as the mass in $Q_m$, namely $m_0$.
The result (\ref{eq:PQapproach})
is very similar to the equation used in Ref.~\cite{mobilityedge} 
to investigate $\rho_Q(\lambda)$ numerically.

To use eq.~(\ref{eq:PQapproach}) one needs to determine the PQ twisted condensate
for real twisted mass
and analytically continue to complex values. 
For an unquenched theory, and working at leading-order in \chpt,
this would simply require minimizing the potential for real masses
and analytically continuing the resulting expression. This minimization
has been worked out in Refs.~\cite{Munster,Scorzato,ShWuphase,AokiBartm},
and analytic continuation is straightforward as the equation to be solved is
a quartic polynomial.
For a PQ theory, however, the methodology is less straightforward.
If one uses the graded-symmetry method~\cite{BGPQ}, the complications are largely
due to the presence of ghosts. The similar problem of studying the phase structure
in {\em quenched} QCD with Wilson fermions was considered in Ref.~\cite{GSS}, and
it was only after making a number of assumptions beyond those of \chpt\
that a result could be obtained. 

Because of this complication, I decided to study
eq.~(\ref{eq:PQapproach}) using the replica approach to
PQ\chpt~\cite{replica}.
This replaces the complication of ghosts with the need
to analytically continue in the number of valence quarks, $N_V$.
This continuation becomes  non-trivial if there is phase structure
which depends on $N_V$, as might be the case here.
My analysis of these issues is preliminary, and I have only
pursued the method far enough to determine the assumptions necessary
to reproduce the results of the primary method, and to obtain
some hints for possible problems with that method. 
Because this alternate analysis is incomplete, I relegate
it to appendix~\ref{app:replica}.

\bigskip

The primary method I use makes use of the formulation given in Ref.~\cite{DD}. 
In this one considers the spectrum of $Q_m^2$, the square of the
Hermitian Wilson-Dirac operator, whose spectral density is 
denoted $\rho(\alpha)$ (without a subscript).
The eigenvalues of $Q_m^2$ are both real and positive, 
with $\bar\alpha$ the smallest. 
To get a feel for $\rho(\alpha)$, it is useful to see the form
it attains in the continuum limit.
Then $Q_m^2 = -\Dslash^2 + m^2$, so the  spectral gap is $m^2$, 
and the spectral density is given in terms of that of $-i\Dslash$ by
\begin{equation}
\rho(\alpha)^{\rm cont} = 
\left\{
\begin{array}{lr}\frac{\rho_{\Dslash}(\sqrt{\alpha-m^2})}
{\sqrt{\alpha-m^2}}\qquad\qquad &\alpha\ge m^2 \\
0 & 0< \alpha < m^2.\end{array}\right.\,. 
\end{equation}
Here I have used the results that the spectrum of $-i\Dslash$ is
symmetric under $\lambda \leftrightarrow -\lambda$, and that exact
zero modes give vanishing contributions in infinite volume.
Using the Banks-Casher relation, eq.~(\ref{eq:rhofromBC}),
one sees that $\rho(\alpha)^{\rm cont}$
has an integrable square-root singularity at the position of the gap
\begin{equation}
\rho(\alpha)^{\rm cont} \approx
-\frac{\langle \bar q_S q_S\rangle}{\pi \sqrt{\alpha-m^2}} 
+ O(\Lambda_{\rm QCD}^2)\qquad\qquad
\alpha\ge m^2 \,.
\label{eq:rhocont}
\end{equation}
Note the first term on the right-hand-side (which is the LO contribution
in \chpt)
dominates for eigenvalues satisfying
$m^2\le \alpha \ll \Lambda_{\rm QCD}^2$, since for them 
$\rho(\alpha) \gg \Lambda_{\rm QCD}^2$.
It is for such eigenvalues that the discussion in this paper applies.

In order to discuss the
renormalization of the spectral density, Ref.~\cite{DD} introduces a resolvent
\begin{equation}
R(z) = \int_{\bar\alpha}^\infty d\alpha 
\,\frac{\rho(\alpha)}{\alpha^2 (z-\alpha)}\,.
\label{eq:resolvent}
\end{equation}
The factor of $\alpha^{-2}$
insures convergence in the ultra-violet (where $\rho$ approaches
its free field form). Thus $R(z)$ defines an analytic function of $z$ with a cut
along the real axis beginning at $\bar\alpha$, across which there is a discontinuity
\begin{equation}
{\rm Disc}[R]\bigg|_{z=\alpha}  = \frac{-2 i \pi \rho(\alpha)}{\alpha^2}
\,.
\label{eq:disc}
\end{equation}
Assuming a non-vanishing gap, i.e. $\bar\alpha>0$, 
$R$ can be expanded about $z=0$:
\begin{equation}
R(z) = \sum_{k=0}^\infty z^k M_k\,,\qquad\qquad
M_k =  -\int_{\bar\alpha}^\infty d\alpha \frac{\rho(\alpha)}{\alpha^{3+k}}
\,.
\label{eq:Rexp}
\end{equation}
This expansion is convergent for $|z|<\bar\alpha$.
It is useful thanks to the fact that each term in the sum can be written
as a partially quenched expectation value~\cite{DD}
\begin{equation}
M_k = 
\sum_{x_1,\dots,x_{n-1}}
a^{4n-4} \langle P_{12}(x_1) P_{23}(x_2) \dots P_{n-1 n}(x_{n-1}) P_{n 1} (0) \rangle_{PQ}
\equiv \langle P^n\rangle_{PQ} \,,
\qquad n=2k+6
\,.
\label{eq:MkPQ}
\end{equation}
The notation here is as follows. There are at least $n$ flavors of valence quarks,
and $P_{j k}$ is a local bare flavor non-singlet pseudoscalar density:
$P_{j k}(x) = \bar q_j \gamma_5 q_{k}(x)$. The flavors are all degenerate,
with the same bare mass, $m_0$, as the sea-quarks of interest.
The $x_i$ represent lattice sites, and
are summed over, so that all insertions are at zero 4-momentum. 
The choice of flavor indices implies that the quark contractions form a single
cycle, so that, upon insertion of the eigenvalue expansion of the quark propagator,
one recovers the form of $M_k$ in eq.~(\ref{eq:Rexp}).
The shorthand notation $\langle P^n\rangle_{PQ}$ is for later convenience.

In order that the spectral density has a well-defined continuum limit,
it must be renormalized. This is accomplished straightforwardly by
renormalizing the coefficients in the expansion of the
resolvent~\cite{DD}:
\begin{equation}
R_R(z) = \sum_{k=0}^\infty z^k M_{k,R}\,,\qquad\qquad
M_{k,R} = Z_P^{2k+6} M_{k} \equiv \langle P_R^{2k+6}\rangle_{PQ} \,.
\label{eq:ressumren}
\end{equation}
Here  $Z_P$ is the renormalization factor for the non-singlet pseudoscalar density
(defined in a convenient scheme), and $P_R = Z_P P$.
It then follows that the renormalized resolvent is given in terms
of the renormalized spectral density by an equation of the same form
as eq.~(\ref{eq:resolvent}), 
and has a discontinuity of the same form as eq.~(\ref{eq:disc}),
but with $\rho(\alpha)\to\rho_R(\alpha)$, where
\begin{equation}
\rho_R(\alpha) = Z_P^2\ \rho(Z_P^2\alpha) \,.
\label{eq:rhoren}
\end{equation}
The point of this result is that it shows how to rescale the
argument and normalization of the lattice spectral density $\rho(\alpha)$ 
in order to obtain a quantity which has a good continuum limit.
In particular, when $a\to0$, $\rho_R(\alpha)$ becomes the
$\rho(\alpha)^{\rm cont}$ discussed above.
This completes the review of the results of Ref.~\cite{DD}.

\bigskip

The strategy of my primary method is straightforward.
First, calculate the correlation functions $\langle P_R^n\rangle_{PQ}$
using PQ\chpt\ for Wilson fermions (PQW\chpt),
and thus obtain the coefficients $M_{k,R}$.
Second, sum the series in eq.~(\ref{eq:ressumren}) 
within its domain of convergence.
Third, analytically continue the result and determine the discontinuity across
the real axis, thus obtaining the renormalized spectral density and its gap.
One can relate the result back to the 
bare lattice spectral density and gap using eq.~(\ref{eq:rhoren}).

In practice, one can work only to a given order in chiral perturbation theory.
Here I work only at leading-order---tree-level---which is simple enough that
one can sum the series. As I show, this order reproduces the Banks-Casher result
in the continuum, i.e. it gives the dominant contribution for low eigenvalues.
The new feature here is the extension of this result
by the inclusion of discretization effects.

In order to facilitate the calculation, it is useful to introduce an
auxiliary function,
\begin{equation}
F(z) = \sum_{\ell=1}^\infty  \frac{z^\ell}{\ell} \langle P_R^{2\ell}\rangle_{PQ}
\,,
\label{eq:fdef}
\end{equation} 
which includes contributions from $\langle P_R^n\rangle_{PQ}$ starting at $n=2$
rather than $n=6$,\footnote{%
Strictly speaking, to renormalize $F(z)$
one needs to subtract the mixing of
$\langle P_R^2\rangle_{PQ}$ and $\langle P_R^4\rangle_{PQ}$ 
with the identity operator, mixing that is
quadratically and logarithmically divergent, respectively.
This mixing appears in \chpt\ through contact terms between sources.
In practice, one can ignore this subtlety (by dropping these
contact terms), as the $\ell=1,2$ terms in
the sum play no role in the development of the non-analyticity that
is the focus here.}
and in which there is an additional power of $\ell$ in the denominator.
Given $F(z)$, the resolvent can be obtained from
\begin{equation}
R_R(z) = \frac{F'(z) - F'(0) - z F''(0)}{z^2}
\,.
\label{eq:Rfromf}
\end{equation}
Thus the definition of the resolvent, eq.~(\ref{eq:resolvent}),
can be viewed as a twice subtracted dispersion relation for $F'(z)$.
It follows from eq.~(\ref{eq:Rfromf}) and (\ref{eq:disc}) that
$F'(z)$ has a cut along the real axis in the same position as $R_R(z)$,
but with discontinuity
\begin{equation}
{\rm Disc}[F']\bigg|_{z=\alpha} = -2 i \pi \rho_R(\alpha)
\,.
\label{eq:discfprime}
\end{equation}
$F(z)$ can be obtained from $F'$ by integrating radially outward from the
origin. It follows that $F$ is analytic
except along the same cut on the real axis, with its
discontinuity being
\begin{equation}
{\rm Disc}[F]\bigg|_{z=\alpha} = -2 i \pi N_R(\alpha) \equiv 
- 2 i \pi \int_{\bar\alpha}^\alpha d\alpha' \rho_R(\alpha)
\,,
\label{eq:discF}
\end{equation}
as long as the integral exists, as will be the case here.
The quantity $N_R(\alpha)$ is the integrated spectral density:
the number of eigenvalues per unit volume
with eigenvalue less than $\alpha$.
Thus, if one can calculate $F(z)$ by summing the series
in eq.~(\ref{eq:fdef}), and analytically continue to determine its
cut, one directly obtains $N_R(\alpha)$.

\section{Partially Quenched Chiral Perturbation Theory for Wilson fermions}\label{sec:chpt}

In this section I review the ingredients needed
for a calculation of the correlation functions $\langle P_R^n\rangle_{PQ}$
using PQW\chpt.\footnote{%
I assume that PQ\chpt\ is a valid effective theory describing PQ simulations.
This issue has been discussed in Ref.~\cite{ShShPhi0}.}
I work in the regime in which
the expansion parameters $p^2/\Lambda_\chi^2$ (with $\Lambda_\chi=4\pi f_\pi$), 
$m/\Lambda_{\rm QCD}$ (with $m$ a renormalized valence or sea-quark mass)
and $a^2\Lambda_{\rm QCD}^2$ are all comparable.
I assume these quantities are small compared to unity,
so that it is reasonable to keep only the leading term in the joint
chiral-continuum expansion.

I use the graded-symmetry formulation of PQ\chpt, although,
as will become apparent, the calculation is step-by-step equivalent to
that using the replica trick.
The partially quenched chiral Lagrangian including 
discretization effects up to $O(a^2)$
has been given in Ref.~\cite{BRS}. 
The LO terms are\footnote{%
The relation to the constants of Ref.~\cite{BRS} is as follows:
$W''_6=W'_6 - W_6 + L_6$, $W''_7=W'_7 - W_7 + L_7$
and $W''_8=W'_8 - W_8 + L_8$.}
\begin{eqnarray}
\CL_\chi &=& \frac{f^2}{4} \Str\left(\partial_\mu \Sigma\partial_\mu \Sigma^\dagger\right)
- \frac{f^2}{4} \Str\left(\chi'_+\Sigma^\dagger+\Sigma\chi'_-\right) \nonumber \\
&&\mbox{}
- \hat{a}^2 W''_6 \ \left[\Str\left(\Sigma+\Sigma^\dagger\right)\right]^2
- \hat{a}^2 W''_7 \ \left[\Str\left(\Sigma-\Sigma^\dagger\right)\right]^2
- \hat{a}^2 W''_8 \ \Str\left(\Sigma^2 + [\Sigma^\dagger]^2\right)
\label{eq:LPQ}
\end{eqnarray}
Here the field $\Sigma\in SU(N_S+N_V|N_V)$ contains the pseudo-Goldstone bosons and
fermions of the graded-symmetry group 
$SU(N_S+N_V|N_V)_L\times SU(N_S+N_V|N_V)_R$~\cite{BGPQ,ShShPhi0},
and ``$\Str$'' indicates supertrace. 
The LECs which appear are as follows:
$f$, the decay constant in the chiral limit, normalized so that $f_\pi=93\;$MeV;
$W_0 \sim \Lambda_{\rm QCD}^3$, which appears in the definition $\hat{a}\equiv 2W_0 a$, 
and represents the generic size of discretization effects;
and the dimensionless constants $W''_{6-8}$, which describe variations
from this generic size.\footnote{%
For my present purposes, this notation is redundant, as $W_0$ can clearly be
absorbed into $W''_{6-8}$. I use this notation, however, in order to be consistent
with Ref.~\cite{BRS}.}
%

The quantities $\chi'_\pm$ in eq.~(\ref{eq:LPQ}) contain the 
renormalized quark mass
matrix and the (matrix) source for the renormalized pseudoscalar density:
\begin{equation}
\chi'_\pm = 2 B_0 (M' \pm p) \,,
\label{eq:chi}
\end{equation}
Here $B_0$ is the standard continuum LEC.
The primes on $\chi'$ and $M'$ follow the notation of Ref.~\cite{tmNLO} and indicate
that the $O(a)$ term has been absorbed by a shift of the quark mass.
The expressions (\ref{eq:chi}) are somewhat unconventional---usually
one has $\chi'_+=2 B_0 (s + ip)$ and $\chi'_-=(\chi'_+)^\dagger$,
with $s$ and $p$ hermitian matrix sources, the former ultimately set to $M'$.
This corresponds at the quark-level in the continuum effective
Lagrangian to using
%
$\CL_q= \bar q(s + i\gamma_5 p) q$, which is the natural choice of the
source in an operator treatment since it is hermitian.
What is of interest here, however, are matrix elements of 
$\bar q_j \gamma_5 q_k$
without the factor of $i$.
These are obtained if the source term in the continuum quark-level Lagrangian has the form
$\CL_q = \bar q(M + \gamma_5 p) q$ (with $p$ not necessarily hermitian).
A standard spurion analysis then
leads to the result of eq.~(\ref{eq:chi}).
No problems are caused by having a complex action
since the sources are being used as a tool to develop perturbation theory.
Note that one obtains the desired pseudoscalar density using
\begin{equation}
P_{jk,R}(x) = \frac{\delta}{\delta p_{jk}(x)} \CS
= - \frac{f^2 B_0}{2} (\Sigma^\dagger-\Sigma)_{kj}
\,,
\label{eq:Pchi}
\end{equation}
where $\CS$ is the action. This is the same form as
in the continuum---discretization errors lead to
corrections suppressed by powers of $a$.
As in the continuum, one automatically obtains at LO
the {\em renormalized} pseudoscalar density from the \chpt\ calculation, 
with the scheme dependence 
implicitly contained in that of the constant $B_0$.

The quark mass matrix is given by
\begin{equation}
M' = {\rm diag}(m_u,m_d,m_s,\underbrace{m_V,m_V,\dots}_{N_V terms},
\underbrace{m_V,m_V,\dots}_{N_V terms})
\,,
\label{eq:massmatrix}
\end{equation}
where I have specialized to $N_S=3$ light sea-quarks.
All masses are renormalized, and are
related to the corresponding bare quark
masses by an additive shift of size $1/a$, an overall renormalization,
and the further additive shift of $O(a)$ mentioned above~\cite{ShSi}.
To obtain the spectral density for one of the hermitian Wilson-Dirac
operators entering into the determinant $m_V$ must equal one
of the sea-quark masses.\footnote{%
This follows from the discussion after eq.~(\ref{eq:MkPQ}).
It is of interest also to consider $m_V\ne m_S$, with $m_S$ a generic
sea-quark mass. This tells one about
the spectral density for partially quenched quarks. I have not considered
this possibility in this work.}
Since it is the light quark determinant that has
the smallest gap, and since almost all simulations to date work in the 
isospin-symmetric limit (``2+1 flavors''),
I choose $m_V=m_u=m_d \ll m_s$.
I imagine the strange quark mass held fixed (at or near its physical value),
while the common light quark mass is extrapolated towards its physical value.
In fact, this choice allows the strange quark to be integrated out of the
chiral theory, avoiding concerns about the convergence of \chpt\ if it is kept.
In the following I assume that this has been done. Thus I will
be considering the effective theory with 2 light degenerate sea-quarks,
for which the effective Lagrangian has
exactly the same form as eq.~(\ref{eq:LPQ}), except
that (i) all matrices lose one row and column, 
(ii) $m_s$ does not appear in $M'$, and 
(iii) all the LECs have an implicit, unknown dependence on $m_s$.
Since $m_s$ is ultimately to be fixed to its physical value, and since
the constants are unknown, the dependence on $m_s$ does not add further 
uncertainty.
Note that the subsequent considerations will also apply directly for
a two-flavor theory (i.e. one in which the strange quark is quenched)---the only
change being that the LECs will have different values.

This completes the description of the PQ chiral theory. Before proceeding
to the calculation, I make some general comments on the set-up.
First, the result that the LO term linear in $a$ can be absorbed 
into the quark mass
immediately implies that the discretization errors in
the spectral density will be of $O(a^2)$.
Note that this does not mean there will be no errors linear in $a$: such errors
are present, but of next-to-leading-order (NLO),
 because they are multiplied by $m$ or $p^2$.

Second, the LECs, and in particular $W''_{6-8}$, are {\em independent of $N_V$}.
This is clear
for those linear combinations 
that remain when one considers the unquenched $SU(2)$ 
subgroup (discussed in more detail below). The effective theory in this subgroup
is independent of $N_V$~\cite{BGPQ,ShSh,ShShPhi0}.
For other linear combinations, which contribute only to PQ correlation functions,
the argument is more indirect. These combinations give the LO contribution to
particular amputated PQ correlation functions 
(which can be chosen so as to cancel the
contributions of the physical LECs, along the lines followed in Ref.~\cite{ShVdW}).
At the quark-level, these correlation functions are independent of $N_V$, 
by construction. 
Since one can, in principle, 
make LO \chpt\ be arbitrarily accurate by sending $m,a^2\to0$,
the ``unphysical'' combinations of LECs must also be independent of $N_V$.
One can also understand this {\em a posteriori}: the spectrum of $Q_m$,
which knows nothing about $N_V$, depends on these combinations.

The third general comment concerns the relative magnitude of $W''_{6-8}$.
It is straightforward to show that, for $N_c$ (the number of colors) large,
$W''_6/W''_8 \sim W''_7/W''_8 \sim 1/N_c$. Thus one expects, even for $N_c=3$,
that the dominant LEC describing discretization errors will be $W''_8$.
I will make use of this expectation when describing  the implications of 
subsequent results, although the results themselves will be general.

Finally, I note that the partial quenching needed for this calculation is,
in some sense, ``minimal''. For one thing, since $m_V=m_S$, there are
no double-pole contributions to propagators. While these would not appear until NLO
in the present calculation, their absence removes the most unphysical
aspect of PQ\chpt. In addition, in the LO calculation I carry out below, the
ghosts (or replicas if using the replica trick) play no role.
The information that PQ\chpt\ is ``supplying'' is that (a) the mass of 
pseudo-Goldstone bosons (PGBs) 
composed of valence quark and antiquark is the same as that of PGBs
composed of sea-quarks [which follows from the exact $SU(N_S+N_V)$ vector symmetry,
assuming this is unbroken] and (b) that one can separate the contributions
from individual contractions in an effective field theory framework.
One cannot avoid partial quenching, however, because
it is not possible with sea-quark correlators alone to pick out
the quark-connected contractions that contribute to
$\langle P_R^n\rangle_{PQ}$.

\bigskip
I close this section by recalling the essentials of the phase structure
in the two flavor theory caused by the $a^2$ terms.\footnote{%
The effects of higher
order terms have been considered in Ref.~\cite{AokiBar2,observations}.}
The phase structure is determined by minimizing the potential in the
physical sub-sector of the theory, i.e. that involving only the two sea-quarks.
Restricting $\Sigma$ to this sector, and setting the source $p$ to zero,
one finds
\begin{equation}
\CV_{UQ} = - \frac{f^2 B_0 m}{2} \Tr(\Sigma^\dagger+\Sigma)
- \hat{a}^2 W' \left[\Tr(\Sigma^\dagger+\Sigma)\right]^2 
\,.
\end{equation}
Here I have used the fact that $\Sigma\in SU(2)$ to simplify terms,
defined $W'=W''_6+W''_8/2$ (following Ref.~\cite{tmNLO}), 
and dropped an irrelevant constant.

At this stage it is useful to introduce a more compact notation.
Extending the notation of Ref.~\cite{observations}, I define 
\begin{equation}
\hat m = 2 B_0 m\,,\quad
w_6 = \frac{16 \hat{a}^2 W''_6}{f^2}\,,\quad
w_8 = \frac{16 \hat{a}^2 W''_8}{f^2}\,,\quad
w'= \frac{16 \hat{a}^2 W'}{f^2}= w_6 + \frac{w_8}{2}\,.
\end{equation}
The phase structure depends on the sign of $w'$~\cite{ShSi}.
(The relation to the notation of Ref.~\cite{ShSi} is $w'= -f^2 c_2$.)
For $w'<0$ there is an Aoki phase, running between end points
at $\hat{m} = \pm 2 |w'|$. The presence of a non-zero twisted
condensate inside the Aoki phase implies that the gap in the
spectrum of $Q_m$ vanishes~\cite{ShSi}. Thus the methodology
developed here applies only away from the Aoki phase.
In this region, the pion mass is given by
\begin{equation}
m_\pi^2 = |\hat{m}| + 2 w'
\,,
\label{eq:mpisq}
\end{equation}
which vanishes at the end points.

For $w'>0$ there is a first-order transition at $m=0$.
The result (\ref{eq:mpisq}) still holds, showing that there
is a minimum pion mass proportional to $a$. The twisted condensate
vanishes along the entire Wilson axis, implying that the gap is non-zero
for all $m$.

\section{Derivation of Primary Method}
\label{sec:chptcalc}

The aim in this section is to use PQW\chpt\ to obtain an explicit
expression for the
auxiliary function $F(z)$, defined in eq.~(\ref{eq:fdef}).
To do so, one needs to determine $\langle P^{n}_R\rangle_{PQ}$ in
a sufficiently explicit form that one can sum the series defining $F$.
I will be able to do this at LO in the chiral expansion, with the
result, after a rather extended saga, being given in eq.~(\ref{eq:final}).

The first step is to map the lattice correlation function into
one in the continuum effective theory. The former involves products
of sums of the form $a^4 \sum_x P_{ij,R}(x)$. These map into
integrals $\int d^4x\; P_{ij,R}(x)$, up to corrections which I
will now argue are suppressed by powers of $a$,
 and thus are of higher order.
Corrections arise from three sources: (I) in the mapping of the
local operators from the lattice to continuum effective theory
(i.e. the usual $O(a)$ corrections to local operators 
in the Symanzik improvement program); (II) in
the mapping of sums into integrals; and (III) from a mismatch
between contact terms in the lattice and continuum theories.
(I) and (II) lead to corrections to the local continuum operators
of the form
$P_R(x)\to P_R(x) [1 + O(a m) ] + a^2 O_5(x)$ where $O_5$ is
a dimension 5 operator transforming as a pseudoscalar 
(e.g. $\bar q \gamma_5 D^2 q$). Since there is no chiral suppression
of the coupling of the pseudoscalar density to pions, and correspondingly
no enhancement of the coupling of $O_5$, the corrections
are suppressed by the explicit factors of $a$.
The third source is more subtle, but can be seen to be suppressed
by $O(a m_\pi^2)$.\footnote{%
Because of the flavor indices, the operator product expansion
in the continuum is schematically $P(x) P(0)\sim S(x)/x^3$. Integrating
over a cell of size $a^4$ gives $\int P(x) P(0) \sim a S(0)$.
In the lattice theory one also has that $a^4 P(0) P(0) \sim a S(0)$, but the coefficient
of $a$ is different. Thus the difference between lattice and
continuum contact terms is also $\sim a S(0)$. The matrix
elements of $a S(0)$ are suppressed relative to those of $\int P(x) P(0)$ by $a m_\pi^2$,
with the $m_\pi^2$ arising from the loss of one pion propagator.
Contributions from the regions where three or more pseudoscalars come into contact
are even more suppressed.
I thank Martin L\"uscher for pointing out the presence of the
contributions from contact terms.}
The conclusion is that, at LO, I can use the naive transcription 
\begin{equation}
\langle P^{n}_R\rangle_{PQ} 
=
\int d^4x_1 \dots d^4x_{n-1}
\langle P_{12,R}(x_1) 
\dots P_{n-1 n,R}(x_{n-1}) P_{n 1,R} (0) 
\rangle_{PQ,{\rm cont}}\,,
\label{eq:lattocontPn}
\end{equation}
and calculate the right-hand-side in PQW\chpt. I will drop
the subscript ``cont'', as the subsequent
calculation  is entirely in the continuum.

From now on, without loss of generality, I
take the quark mass to be positive. It must be large enough to avoid
a possible Aoki phase (where we know that the gap closes and the method breaks down).
The condensate in the sea sector (which I will also refer to as the unquenched [UQ]
sector)
is then aligned in the standard way,
$\langle\Sigma_{UQ}\rangle = 1$, and, in particular, parity and flavor are 
not spontaneously broken. Assuming the graded vector symmetries are unbroken 
this further implies that $\langle \Sigma_{PQ}\rangle = 1$,
i.e. that the partially quenched condensate is unity~\cite{ShShPhi0}.
One then expands in pseudo-Goldstone bosons and fermions in the usual way,
$\Sigma = \exp(2 i\Phi/f)$. Since parity and flavor symmetries are unbroken,
$\langle \Phi^n\rangle=0$ for $n$ odd.

\begin{figure}[t]
\begin{center}
\epsfysize=1.4truein 
\epsfbox{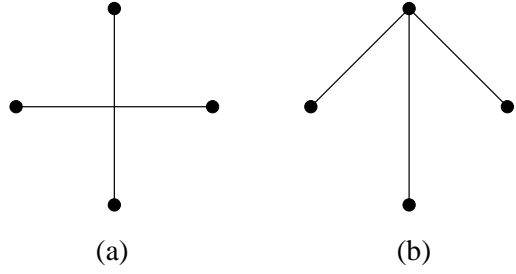}
\end{center}
\caption{\footnotesize
Diagrams contributing to $\langle P^n_R\rangle_{PQ}$ 
with $n=4$ at tree-level
in chiral perturbation theory. Filled circles represent the
pseudoscalar operators.}
\label{fig:4pt}
\end{figure}

\begin{figure}
\begin{center}
\epsfysize=1.4truein 
\epsfbox{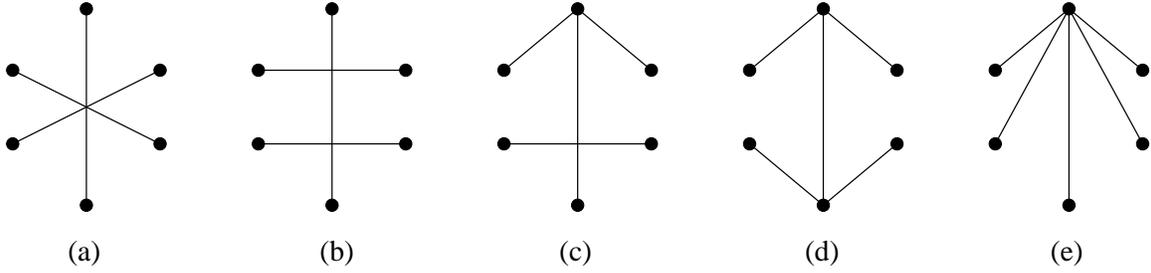}
\end{center}
\caption{\footnotesize
Diagrams contributing to $\langle P^n_R\rangle_{PQ}$ 
with $n=6$ at tree-level
in chiral perturbation theory.}
\label{fig:6pt}
\end{figure}

It is useful to look at low-order examples to understand the
structure of the contributions.
For brevity, I refer to all pseudo-Goldstone particles as pions.
The diagrams contributing at tree-level
to $\langle P^{n}_R\rangle_{PQ}$ for $n=4$ and $6$
are shown in Figs.~\ref{fig:4pt} and \ref{fig:6pt}, respectively.
The cyclic arrangement of flavor indices in eq.~(\ref{eq:MkPQ})
implies that there are no disconnected contractions.
Since all external 4-momenta vanish, the propagators are simply $1/m_\pi^2$
[with $m_\pi^2$ given in eq.~(\ref{eq:mpisq})], while the vertices arise from
the terms in the LO potential [all but the first term in eq.~(\ref{eq:LPQ})].
Note that the pseudoscalar operator, given in eq.~(\ref{eq:Pchi}), can
produce any odd number of particles.

\begin{figure}[t]
\begin{center}
\epsfxsize=\hsize
\epsfbox{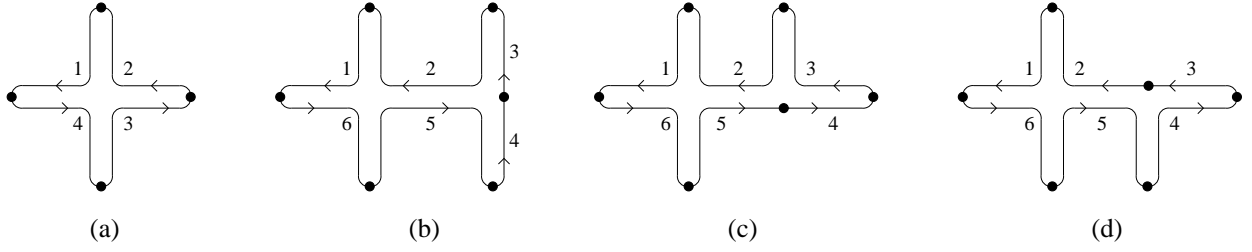}
\end{center}
\caption{\footnotesize
Examples of allowed quark-flow diagrams,
with filled circles representing the
pseudoscalar fields.
(a) Quark-flow diagram corresponding to
fig.~\protect\ref{fig:4pt}a.
(b-d) The three types of quark-flow diagram corresponding
to fig.~\protect\ref{fig:6pt}c. Examples of flavor
labels are shown; they can also
be cyclically permuted ($1\to2$, $2\to3$, etc.).}
\label{fig:4ptqline}
\end{figure}

\begin{figure}
\begin{center}
\epsfysize=1.5truein 
\epsfbox{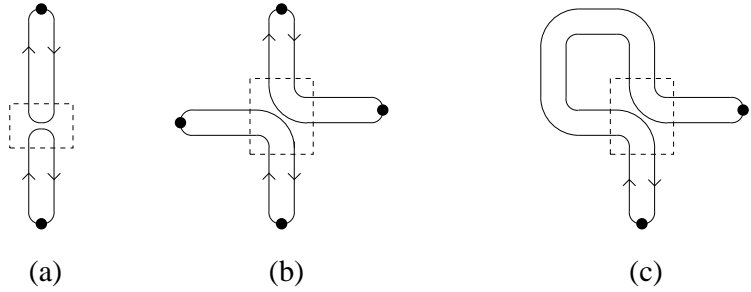}
\end{center}
\caption{\footnotesize
Examples of quark-flow diagrams containing
vertices with two-supertrace operators, in which
both supertraces are contracted with pion fields. 
Filled circles
represent pseudoscalar fields and dashed boxes the vertices.
(a) Disallowed contribution to $\langle P_R^2\rangle_{PQ}$.
(b) Disallowed contribution to $\langle P_R^4\rangle_{PQ}$
corresponding to the diagram in
fig.~\protect\ref{fig:4pt}(a).
(c) Allowed one-loop contribution to 
$\langle P_R^2\rangle_{PQ}$.}
\label{fig:twostrace}
\end{figure}

The cyclic arrangement of flavor indices impacts the calculation
in two other ways. First, it constrains the contribution
from the two-supertrace operators.
In order to get a non-vanishing result, each vertex can only contain a
single-supertrace, as illustrated by the examples in fig.~\ref{fig:4ptqline}.
This happens automatically for the single-supertrace terms
(those proportional to $\chi'_\pm$ and $W''_8$),
but not for those with two supertraces
(those proportional to $W''_6$ and $W''_7$). 
For the latter, quark-level
contractions such as those in fig.~\ref{fig:twostrace}(a-b)
are not allowed by the external flavor indices. 
Only one of the two supertraces in each such vertex
can be contracted with pions; in the other supertrace,
$\Sigma$ and $\Sigma^\dagger$ must be replaced by unity.
This implies that the $W''_7$ term does
not contribute, and that the effect of $W''_6$ can be included by shifting
the mass term. 
In particular, to calculate
$\langle P^{n}_R\rangle_{PQ}$ one can replace $\CL_\chi$ with
\begin{equation}
\CV^{\aux}_\chi = - f^2
\left\{
\frac{\Str\left([\hat{m}^\eff+ \hat{p}]\;\Sigma^\dagger
                  +\Sigma\;[\hat{m}^\eff-\hat{p}]\right)}{4}
+ w_8\; \frac{\Str\left(\Sigma^2 + [\Sigma^\dagger]^2\right)}{16}
\right\}
\,,\label{eq:Leff'}
\end{equation}
where
\begin{equation}
\hat{m}^\eff = \hat{m} +2 w_6 \,,
\end{equation}
and $\hat{p}=2B_0 p$.
The kinetic term has been excluded because all momenta vanish
in the tree-level computation.
This prescription of replacing $\CL_\chi$ with
$\CV^\aux_\chi$ works for all $n$,
including the two-point pseudoscalar correlation
function (i.e. $n=2$). 

I note in passing that the use of the shifted mass $\hat{m}^\eff$
does not simply generalize beyond tree-level.
At NLO, two-supertrace operators have contributions
in which {\em both} supertraces are connected
to pions. An example for the two-point function
is shown in fig.~\ref{fig:twostrace}(c).

It is useful as a check of
subsequent manipulations to determine the explicit
result for $\langle P_R^n\rangle_{PQ}$ at LO for the simplest cases.
I find
\begin{eqnarray}
\langle P_R^2\rangle_{PQ,LO} &=& - f^2 (B_0)^2 \frac{2}{m_\pi^2} 
\label{eq:P2pert}\\
\langle P_R^4\rangle_{PQ,LO} &=& - f^2 (B_0)^4 \frac{4\;
 \hat{m}^\eff}{(m_\pi^2)^4} 
\label{eq:P4pert}\\
\langle P_R^6\rangle_{PQ,LO} &=& 
- f^2 (B_0)^6 \frac{12 \; \hat{m}^\eff(\hat{m}^\eff - w_8)}{(m_\pi^2)^7}
\,, \label{eq:P6pert}
\end{eqnarray}
with $m_\pi^2= \hat{m}^\eff+  w_8$, consistent with eq.~(\ref{eq:mpisq}).

In the Aoki-phase power counting, 
$\hat{m} \sim \hat{m}^\eff \sim w_8 \sim m_\pi^2$, and it is
easy to see that
\begin{equation}
\langle P_R^n\rangle_{PQ} \sim f^2 (B_0)^n \hat{m}^{1-n}  
[ 1 + O(\hat{m}/\Lambda_\chi^2) ]
\,.
\label{eq:powercount}
\end{equation}
The fact that the order of the infra-red divergence
increases with $n$ shows that the convergence
of the sum defining $F(z)$ [eq.~(\ref{eq:fdef})]
appears to break down when $z\sim m^2$, corresponding to
the appearance of non-analyticity.
Note that the NLO chiral corrections from all terms in
the sum are equally sub-dominant in this regime. 
For example, although the NLO correction to,
say, $\langle P_R^6\rangle_{PQ}$ is of the same absolute size
as the LO contribution to $\langle P_R^4\rangle_{PQ}$, the
overall factors of $z$ make the LO contributions to 
$F(z)$ of these two quantities comparable in the range of interest.
The only exception to the dominance of the LO contributions 
is if they turn out not to give a non-analyticity, while the NLO
terms do. This may in fact be the case here for certain values
of the LECs.

\bigskip

Further analytic progress is facilitated by
observing that the PQ correlation functions can be expressed
in terms of simpler unquenched correlators, if one
works only at LO in the chiral expansion.
Explicitly, I claim that
\begin{equation}
2 (n-1)! \langle P_R^n \rangle_{PQ,LO} = 
\langle(P_{R,12}+P_{R,21})^n\rangle_{CONN,LO}\,,\qquad
(n\ge 1)
\,,
\label{eq:key}
\end{equation}
where I use the shorthand notation
\begin{equation}
\langle(P_{R,12}+P_{R,21})^n\rangle_{CONN,LO}
\equiv
\frac1V
\langle \left\{\int d^4x [P_{R,12}(x)+P_{R,21}(x)]\right\}^n\rangle_{CONN,LO}
\,.
\end{equation}
In words, the new correlation function involves $n$ insertions of
the {\em same} operator. This operator contains only two flavors,
which can be chosen to be the flavors of the sea-quarks since the
masses are the same. Thus it is an unquenched correlation function.
It is to be evaluated at tree-level,
using the auxiliary chiral potential
$\CV^\aux_\chi$ of eq.~(\ref{eq:Leff'}),
keeping only the connected contributions. In this calculation,
one can replace the supertraces with traces, since only two flavors enter.
Finally, note that I have integrated all $n$ of the operators over space-time,
rather than just the first $n-1$, as in eq.~(\ref{eq:lattocontPn}), and compensated by the
overall factor of $1/V$, with $V$ the space-time volume. This makes the
expression more symmetric, which is useful below. The use of a finite,
rather than infinite, volume has no impact on the correlation functions
as long as the box size is much larger than the pion Compton wavelength.
This is because the correlation function is connected, and thus falls off
exponentially as the separation between operators grows. Since the volume
can be taken arbitrarily large, finite volume effects can be made
arbitrarily small.

I give a demonstration of eq.~(\ref{eq:key}) in appendix~\ref{app:key}. Here
I attempt to make the result plausible.
One should keep in mind that it is a useful trick,
and not a fundamental result.
In particular, the equality does not hold either if there are terms 
in the Lagrangian with two (super)traces, or beyond
tree-level.

The result is trivially true for odd $n$, since both sides of (\ref{eq:key}) vanish.
It is also trivially true for $n=2$,
as the two sides are identical, taking into account that, 
on the right-hand side,
only the cross-terms $\langle P_{R,12} P_{R,21}\rangle_{LO}$ contribute,
and there are two of these.
The first non-trivial case is for $n=4$.
At this order there are potential
disconnected contributions to the right-hand side
(schematically of the form $\langle P_{R,12} P_{R,21}\rangle_{LO}^2$),
which have no correspondents on the left-hand side,
but these are to be removed by definition.
As for the connected contributions,
the key point is that, even though the flavor indices on
the right-hand side do not themselves force the contractions to
form a single quark-loop, the fact that the vertices have a single
(super)trace does enforce this. 
For example, the quark-flow diagram for fig.~\ref{fig:4pt}a
is given by that of fig.~\ref{fig:4ptqline}a, with 
flavor labels changed as follows: $3\to 1$ and $4\to 2$.
The lack of two-(super)trace vertices is crucial here:
such vertices would lead to
contractions such as that of 
fig.~\ref{fig:twostrace}b on the right-hand side
of eq.(\ref{eq:key}), but not on the left-hand side.
The counting of contractions is, of course, different
in the two expectation values. On the right-hand side of
(\ref{eq:key}), one has an extra choice not available
on the left-hand side (whether to start with $P_{R,12}$ 
or $P_{R,21}$ in a particular term), leading to the
overall factor of $2$. In addition, the
subsequent choices of which $P$ to contract with are fixed
by the flavor indices on the left-hand side, 
there is no such constraint on
the right-hand side, so that there are (for $n=4$) an
additional $3!$ contractions. This is most clear for
fig.~\ref{fig:4pt}a, but also holds for fig.~\ref{fig:4pt}b.

This argument generalizes straightforwardly for the simplest
diagram at each order, that with a single vertex producing $n$ pions
(e.g. figs.~\ref{fig:4pt}a and \ref{fig:6pt}a): the relative
contraction factor is $2 (n-1)!$.
I show in appendix~\ref{app:key} that this result holds for all diagrams.

Another check on the relative factor 
can be obtained as follows. Assuming that the two sides 
of eq.~(\ref{eq:key}) are proportional
(which is established in the appendix), and that one is working
at small enough $m_\pi^2$ so that the LO result is accurate,
one can calculate the relative size of the two sides
at the quark-level. Since the PQ correlator involves only
a single, cyclic contraction, it must be that taking the
connected part of the unquenched correlator 
(which, recall, is to be done at the pion-level) 
is equivalent to taking the connected part at the quark-level
(again, when working at LO in \chpt). This is because the
quark-connected contractions are identical on the two sides, while
quark-disconnected contractions contributing to the unquenched
correlator have different dependences on the volume.
However, the number of quark-connected contractions 
for the unquenched correlator is $2 (n-1)!$, compared to
only one for the PQ correlator, and so the relative factor
is as claimed.

\bigskip

To proceed it is useful to introduce a further auxiliary function:
\begin{equation}
G(\mu) = \sum_{n=1}^{\infty} \frac{(-i \mu)^n}{n!} 
\langle (P_{R,12}+P_{R,21})^n \rangle_{CONN}
\,.
\label{eq:gdef}
\end{equation}
Both even and odd powers of $n$ are included in
the sum, even though the latter vanish. This implies that
$G$ is really a function of $\mu^2$ rather than $\mu$.
The connected correlation functions in the sum
are to be evaluated with the auxiliary Lagrangian
$\CL^\aux_\chi$ rather than $\CL_\chi$. 
Note that they are not truncated at LO in the definition
of $G$, although in practice I will work only at LO.
For this reason I do not need to specify counterterms.

Given the result (\ref{eq:key}) the functions $G$ and $F$ are
related at LO as follows:
\begin{eqnarray}
G(\mu)_{LO} 
&=& 
\sum_{n=1}^{\infty} \frac{(-i \mu)^n}{n!} 2 (n-1)!
\;\langle P_R^n \rangle_{PQ,LO} \\
&=&
\sum_{n=1}^{\infty} \frac{(-i \mu)^n}{n/2}
\;\langle P_R^n \rangle_{PQ,LO} \\
&=&
\sum_{\ell=1}^{\infty}
\frac{(-\mu^2)^{\ell}}{\ell}
\;\langle P_R^n \rangle_{PQ,LO} \\
&=& F(-\mu^2)_{LO}\,.
\end{eqnarray}
On the penultimate line I have used the fact that terms with odd powers of
$\mu$ vanish. Thus if one can sum the series defining $G(\mu)$ at
LO in the chiral expansion, then, by continuing from positive
to negative $\mu^2$ (i.e. from real to imaginary $\mu$) one can obtain
the desired function $F$ in the regime where its discontinuity
gives the integrated spectral density.

Summation of the series defining $G(\mu)$ is straightforward,
using the fact that the log of the partition function generates
connected correlation functions: 
\begin{eqnarray}
G(\mu) &=& \sum_{n=1}^{\infty} \frac{(-i \mu)^n}{n!} 
\langle (P_{R,12}+P_{R,21})^n \rangle_{CONN} \\
&=& \sum_{n=1}^{\infty} \frac{(-i \mu)^n}{n!} 
\frac{1}{V}
\left[-\int(\frac{\delta}{\delta p_{12}}+\frac{\delta}{\delta p_{21}})\right]^n
\ln Z_{\chi}^\aux(p)\bigg|_{p=0} \\
&=& \frac{1}{V} \ln\left[ Z_{\chi}^{\aux}(\mu)/Z_{\chi}^{\aux}(0)\right]\,,
\end{eqnarray}
where $Z_{\chi}^{\aux}(\mu)$ is the partition function evaluated
using the auxiliary chiral Lagrangian but now with the addition of
a twisted mass
\begin{eqnarray}
 \CL^{\aux}_\chi(\mu) &=&  \frac{f^2}{4} 
\Tr(\partial_\mu\Sigma\partial_\mu\Sigma^\dagger)
+ \CV^\aux(\mu) \label{eq:Lauxmu}\\
\CV^\aux(\mu) &=&
-f^2 \left\{
\frac{\Tr\left([\hat{m}^\eff+ i\hat\mu\tau_1]\;\Sigma^\dagger
                  +\Sigma\;[\hat{m}^\eff-i\hat\mu\tau_1]\right)}{4}
+ w_8\; \frac{\Tr\left(\Sigma^2 + [\Sigma^\dagger]^2\right)}{16}
\right\} \,.
\label{eq:Vauxmu}
\end{eqnarray}
Here $\hat\mu = 2 B_0 \mu$.
Again, counterterms are not shown since I will evaluate this partition
function only at LO. Note that the twist is in the $\tau_1$ direction,
because of the choice of flavor indices in eq.~(\ref{eq:key}). 

The final step uses the result that the LO partition function
is given by the classical result:
\begin{equation}
Z_{\chi}^{\aux}(\mu)_{LO} = 
\exp\left[ -S_{\rm min}(\mu) \right] \,,
\end{equation}
assuming that $\mu$ is real so that the action, $S$, is real.
Thus one obtains
\begin{eqnarray}
F(-\mu^2)_{LO} &=&
G(\mu)_{LO} \\
&=& \frac{1}{V} \ln \left[Z_{\chi}^{\aux}(\mu)/Z_\chi^\aux(0)\right] \\
&=& - \CV^\aux_{\rm min}(\mu) + \CV^\aux_{\rm min}(0)\,.
\label{eq:FtoCV}
\end{eqnarray}
The second term is $\mu$-independent constant which does not
contribute to the discontinuity and will henceforth be dropped.

\bigskip
In summary, the method I arrive at is as follows. First, minimize the
auxiliary unquenched potential $\CV^\aux(\mu)$ of eq.~(\ref{eq:Vauxmu}) 
for real $\mu$. Due to parity, the result in fact depends on $\mu^2$.
Second, analytically continue the result to complex $z=-\mu^2$.
Finally, determine the discontinuity for positive real $z$ (negative $\mu^2$),
which, using eq.~(\ref{eq:discF}), gives the integrated
spectral density and along with this the gap:
\begin{equation}
{\rm Disc}\ [\CV^\aux_\min(\mu^2=-z)]\bigg|_{z=\alpha}
= -2 i \pi N_R(\alpha)
\,.
\label{eq:final}
\end{equation}
I show in appendix~\ref{app:replica} how, under certain
assumptions, one obtains an
equivalent result using the alternate method based on 
eq.~(\ref{eq:PQapproach}).

\section{Results}
\label{sec:results}
In order to test the method, I first check that it reproduces the
LO continuum result. Setting $w_6=w_8=0$, the 
auxiliary potential is identical to that of the unquenched continuum theory
with a twisted mass, i.e.
\begin{equation}
\CV^{\rm cont}(\mu) =
-f^2 
\frac{\Tr\left([\hat{m}+ i\hat\mu\tau_1]\;\Sigma^\dagger
                  +\Sigma\;[\hat{m}-i\hat\mu\tau_1]\right)}{4}
\,.
\end{equation}
This is minimized by 
\begin{equation}
\Sigma_0 = \frac{\hat{m} + i\hat\mu \tau_1}{\sqrt{\hat{m}^2 + \hat{\mu}^2}}
\,,
\end{equation}
so that the minimum of the potential is
\begin{equation}
\CV^{\rm cont}(\mu)_{\rm min} = - f^2 \sqrt{\hat{m}^2 + \hat{\mu}^2}
= - 2 f^2 B_0 \sqrt{m^2 + \mu^2}
\,.
\end{equation}
As claimed above, this is a function of $\mu^2$ which can be analytically
continued to give, up to an irrelevant constant,
\begin{equation}
F(z) = 2 f^2 B_0 \sqrt{m^2 -z}
\,.
\label{eq:contF}
\end{equation}
This function has the expected analytic form, i.e. it is
analytic except for a cut along the real axis, starting at
the gap $\bar\alpha=m^2$, with a discontinuity
$-4 i f^2 B_0 \sqrt{\alpha-m^2}$. Using eq.~(\ref{eq:discF}) one
thus finds that
\begin{equation}
N_R(\alpha) = \frac{2 B_0 f^2}{\pi} \sqrt{\alpha-m^2} \qquad
\Rightarrow\qquad \rho_R(\alpha) = \frac{B_0 f^2}{\pi \sqrt{\alpha-m^2}}
\,,
\label{eq:contNR}
\end{equation}
which agrees with the Banks-Casher result, eq.~(\ref{eq:rhocont}),
since $\langle \bar q q\rangle = - B_0 f^2$ at LO in \chpt.
One can also expand $F$ in powers of $z$ and, using eq.~(\ref{eq:fdef}),
reproduce the results for $\langle P_R^n\rangle_{PQ}$
in eqs.~(\ref{eq:P2pert}-\ref{eq:P6pert}) for $w_8=0$.

\bigskip
Now I apply the method for $a\ne 0$. Mathematically, the auxiliary
potential is identical to that which arises when studying the phase
structure of twisted mass QCD, although the coefficients are different.
This problem has been studied in Refs.~\cite{Munster,Scorzato,ShWuphase,AokiBartm}.
In particular, it is shown in Ref.~\cite{ShWuphase} that the minimum of
the potential (for real $\mu$) occurs when
 the condensate is aligned in the same direction
as the twist in the mass term. In the present case this means that
$\Sigma = \exp(i\omega \tau_1)$ and the auxiliary potential becomes
\begin{equation}
\frac{\CV^\aux(\mu)}{f^2}
= - \hat{m}^\eff \cos\omega - \hat\mu \sin\omega
-\frac{w_8}{4} \cos(2\omega)
\label{eq:Vauxvsomega}
\end{equation}
Extremizing leads to the quartic
\begin{equation}
w_8^2 c^4 + 2 w_8 \hat{m}^\eff c^3 + 
[(\hat{m}^\eff)^2 + \hat{\mu}^2 - w_8^2] c^2 - 2 w_8 \hat{m}^\eff c = (\hat{m}^\eff)^2
\,,\label{eq:quartic}
\end{equation}
where $c\equiv\cos\omega$, and I recall that a ``hat'' implies
multiplication by $2 B_0$. The extrema of the potential are given
in terms of the roots, which I label $\bar c$, by
\begin{equation}
\frac{\CV^\aux(\mu)_{\rm ext}}{f^2}
= - \hat{m}^\eff \bar{c} - \frac{\hat{\mu}^2 \bar{c}}{\hat{m}^\eff + w_8 \bar{c}}
-\frac{w_8}{4} (2\bar{c}^2-1)
\,.
\end{equation}
Note that, as claimed above, both these extrema and the roots themselves
depend on $\mu^2$, despite the fact that the potential itself is linear in $\mu$.

In order to interpret the subsequent results, it is useful to study
the physical meaning of the parameters than can be varied.
Recall that the original PQ chiral Lagrangian depended on two parameters
describing discretization errors ($w_6$ and $w_8$), the quark mass
(conveniently packaged into $\hat{m}=2B_0 m$),
and the decay constant $f$. The latter enters only as an overall factor
and does not play a role in determining the size of the gap or the shape
of the spectral density. 
Different linear combinations of the three other parameters enter
into the physics of the unquenched (sea-quark) theory and the spectral density.
The phase structure and pion properties in the unquenched
theory are determined by $w'=w_6+w_8/2$ and $\hat{m}$, as explained
at the end of sec.~\ref{sec:chpt},
while the spectral density depends on $w_8$ and $\hat{m}^\eff=\hat{m}+2 w_6$.
The situation is potentially confusing because to study the spectral density
one uses the auxiliary Lagrangian, and this is identical in structure
to that describing the unquenched theory, but with different LECs.

\begin{figure}[tb]
\begin{center}
\epsfysize=4.5truein 
\epsfbox{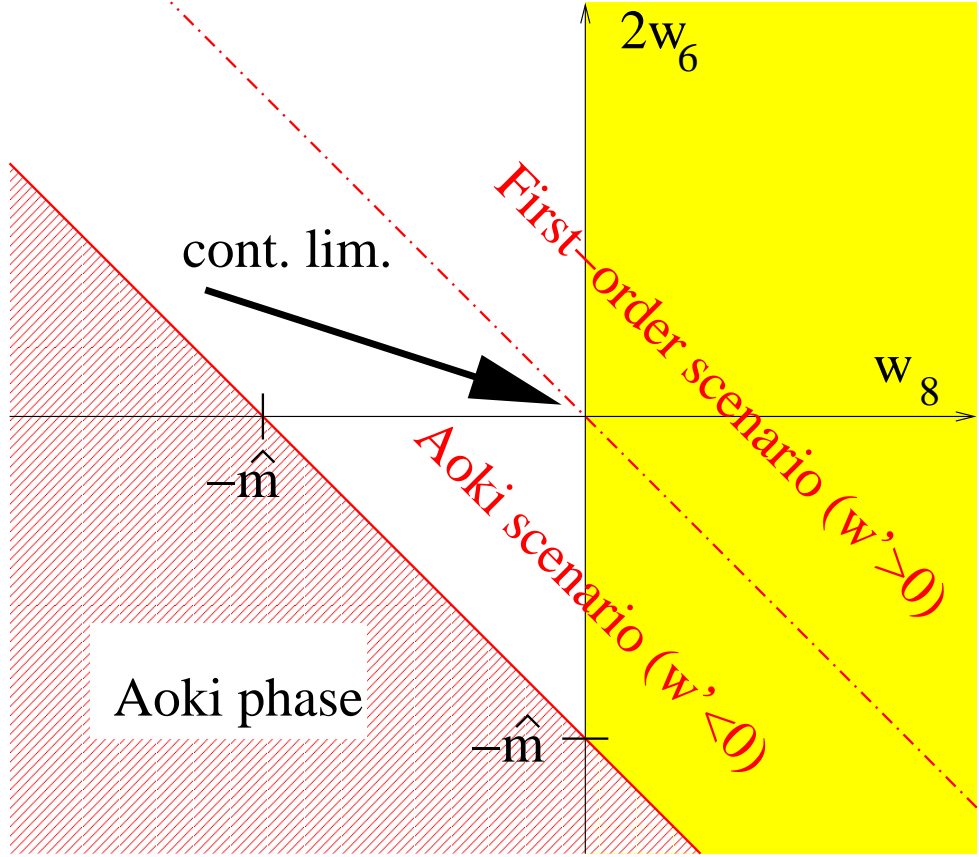}
\end{center}
\caption{\footnotesize
Dependence of phase structure
and spectral density on LECs describing discretization errors, 
for fixed $\hat{m}$, and at LO in \chpt.
The Aoki phase in the unquenched (sea-quark) theory
occurs in the hatched [red] region.
Lines of constant $m_\pi^2$ are parallel to its
boundary, with the dot-dashed line corresponding to 
the continuum result $m_\pi^2=\hat{m}$. This line also marks the
boundary between Aoki-phase and first-order 
scenarios for the unquenched theory. 
For a given gauge and fermion action, the continuum limit
is taken by approaching the origin at fixed $2w_6/w_8$, as in the
example shown by the thick arrow. 
The method used here is inapplicable in the Aoki phase,
and fails for $w_8>0$ (light [yellow] shading). Thus it works
only in the unshaded triangular region.
}
\label{fig:w6w8A}
\end{figure}

I try here to provide a guide to the parameter space.
The dependence of properties of the unquenched theory on the parameters
is shown in fig.~\ref{fig:w6w8A}. The method I use breaks down in
the Aoki phase ($2w'< - \hat{m}$).
If  $-\hat{m}< 2 w' < 0$ one is in the Aoki-phase scenario 
(though away from the Aoki phase itself), while if $0<w'$ one is
in the first-order scenario. If one takes the continuum limit
using a particular choice of gauge and fermion action, and holding
$\hat{m}$ fixed, then $2w_6$ and $w_8$ will both decrease as $a^2$,
but their ratio $r_{68}=2 w_6/w_8$ will be fixed 
up to logarithmic corrections. Thus one approaches the origin
in the figure along a line of slope $r_{68}$ (so that one remains
in one or other of the scenarios).
The issue is what happens to the gap and spectral density as one does so.
The actual value of the slope depends on the actions, but
one expects from large $N_c$ arguments that $|r_{68}|<1$,
as in the example in the figure.


It turns out that the primary method developed in this paper
only yields a prediction if $w_8\le 0$, as will be explained below.
One way to phrase this result is that the method works only
if theory described by the auxiliary Lagrangian 
is in its Aoki-phase scenario. Conversely, the method fails
if the auxiliary theory is in its first-order scenario.
Combining the constraints, 
one finds, as shown in fig.~\ref{fig:w6w8A},
that the method only works in a
triangular wedge of parameter space. 
This allowed wedge contains regions of both 
the Aoki-phase and first-order scenarios of the 
original, unquenched theory.
If one assumes that the LECs take their most likely
values such that $|r_{68}|<1$,
then the allowed wedge contains only the Aoki-phase scenario
of the unquenched theory. 
In other words, if $|r_{68}|<1$, the scenarios in the auxiliary
theory and the original theory match, and if the
former must be in the Aoki-phase scenario for the primary method
to work, so must the latter.

\begin{figure}[tb]
\begin{center}
\epsfysize=4.5truein 
\epsfbox{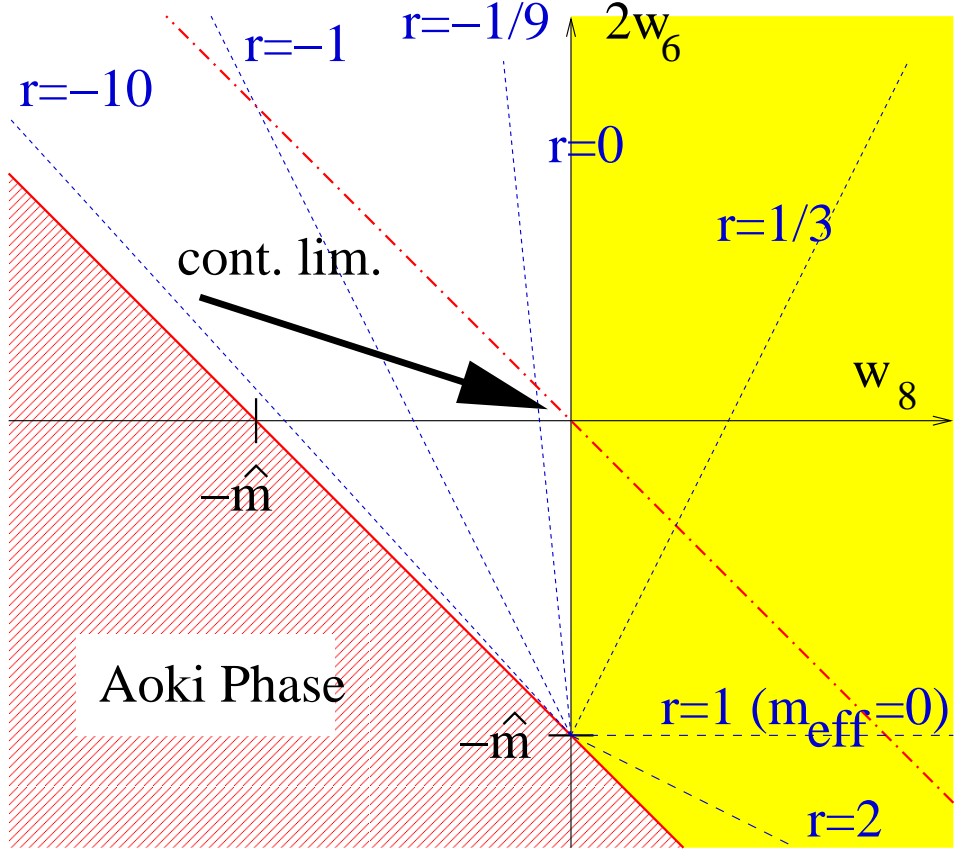}
\end{center}
\caption{\footnotesize
Parameter space (as in fig.~\protect\ref{fig:w6w8A})
showing lines of constant $r$, along which the form of
discretization errors in the spectral density does not change.
Note that the line with $r=1$ also corresponds to $\hat{m}^\eff=0$.
}
\label{fig:w6w8B}
\end{figure}

The roots of eq.~(\ref{eq:quartic}) depend on two dimensionless ratios,
which I choose to be
\begin{equation}
r \equiv \frac{w_8}{m_\pi^2} = \frac{w_8}{\hat{m} + 2 w_6 + w_8}
\qquad
{\rm and}
\qquad
z' \equiv \frac{(2 B_0)^2 z}{m_\pi^4} =
\frac{-\hat{\mu}^2}{m_\pi^4}
\,,
\end{equation}
where $m_\pi^2$ is the value for the unquenched sea-quark pion
(which is unaffected by the twisted valence mass).
Roughly speaking, $r$ characterizes the relative size of the
contributions to $m_\pi^2$ due to discretization errors and the quark mass.
Its varies in magnitude from $0$ in the continuum limit
to $\infty$ at the Aoki phase end points.
Its sign matches that of $w_8$.
Thus for $w_8<0$, its range is $-\infty< r< 0$, 
while for $w_8>0$, the range is $0 < r < +\infty$.
Some lines of constant $r$ are shown in fig.~\ref{fig:w6w8B}.
The importance of these lines is that the distortion
of the spectral density and the gap by discretization effects 
is constant along them.
When taking the continuum limit, one crosses these lines,
and the distortion changes.
The rate of this change depends on the slope of the
approach to the continuum (i.e. on $r_{68}$), but the
qualitative nature of the approach to the continuum is
independent of the slope.
What does matter, however, is whether one approaches from
positive or negative $w_8$.

I first describe the results for $w_8\le 0$, and then explain the
failure of the method for $w_8>0$.
I will show results for $r= -10,\ -1,\ -1/9$ and $-1/99$,
which span the ``allowed'' region of parameter space, as shown
in fig.~\ref{fig:w6w8B} ($r=-1/99$ lies slightly to the left
of the $2w_6$ axis).
This sequence of $r$ values corresponds to
varying the discretization errors from large to very small.

It may be useful to give another
characterization of the size of discretization errors.
This is the ratio, $r'$, of the quark mass to the
half-length of the phase boundary in the unquenched theory, 
$L = |w'|/B_0$. Note that 
the boundary has the same length in both scenarios, and 
is a concrete measure of the size of discretization errors.
This new measure is related to $r$ by
\begin{equation}
r'=\frac{{m}}{L} = \frac{\hat{m}}{|2 w'|}
= \frac{1 - r (1 + r_{68})}{|r (1 +r_{68})|}
\,.
\end{equation}
If $r_{68}=0$, then $r'$ is determined by $r$,
and the choices of $r$ listed above correspond to
$r' =1.1,\ 2,\ 10$ and $100$, respectively. 
In words, these values
correspond to a quark mass very close to the Aoki-phase end point, 
one half-length away, ten half-lengths away and one hundred half-lengths away.
If $r_{68}$ is non-zero, 
the values of $r'$ corresponding to the standard set of $r$ change, 
but the qualitative feature of a transition from
very large to very small discretization errors does not.

\begin{figure}
\centering
\subfigure[]{
\label{fig:reroot1}
\scalebox{1}[1]{\includegraphics[width=3in]{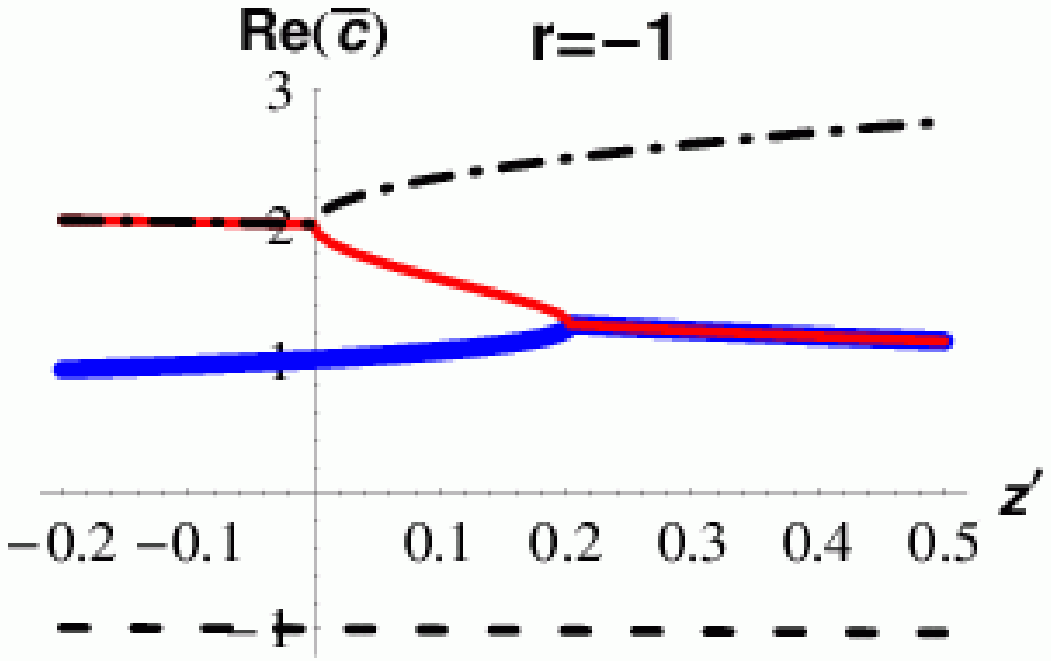}}}
\subfigure[]{
\label{fig:imroot1}
\scalebox{1}[1]{\includegraphics[width=3in]{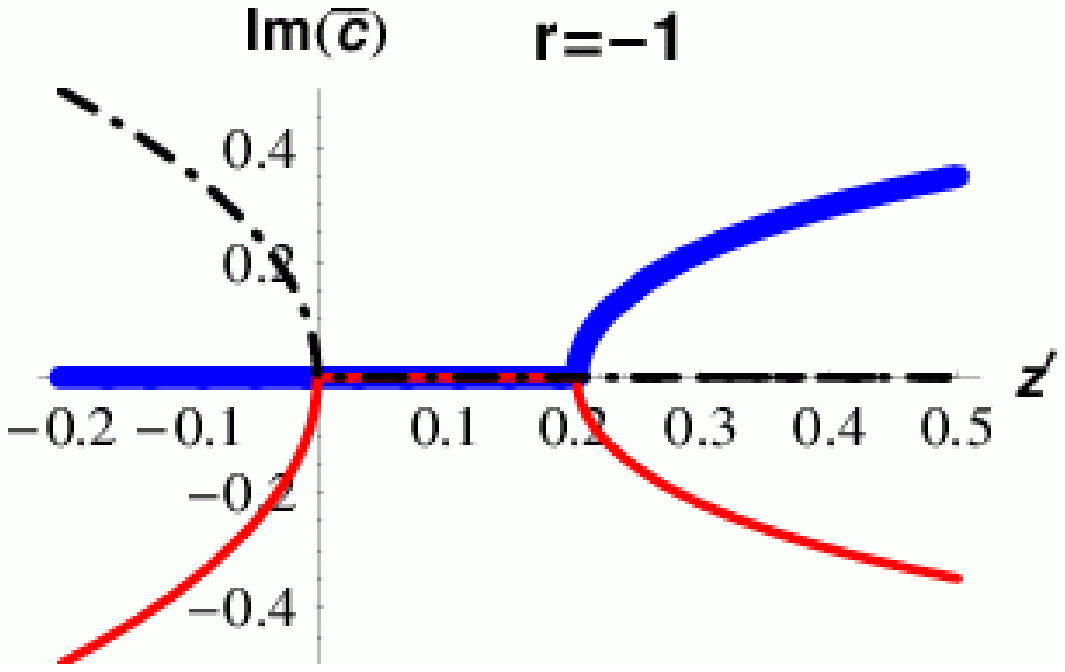}}}
\caption{\label{fig:root1}
\footnotesize
Real and imaginary parts of roots $\bar c$ for $r=-1$,
plotted against $z'$. The variable $z'$ is chosen
so that the continuum gap occurs at $z'=1$. 
The thickest solid curve (blue) shows the root of interest,
which has $\bar c=1$ at $z'=0$. It joins with the thinner solid curve
(red) at the start of the gap, and for large $z'$ these form a complex
conjugate pair. The other roots are shown as dashed and dot-dashed.
}
\end{figure}

I begin by showing results for $r=-1$, the value for which
discretization and quark mass contributions are comparable.
The real and imaginary parts of the roots $\bar c$ are shown
in fig.~\ref{fig:root1}. Since $z'\propto -\mu^2$, negative
$z'$ corresponds to real twisted mass where the potential is
to be minimized. This picks the root with $\bar c=1$ at $z'=0$,
i.e. that for which the condensate is $\Sigma=1$ when the twisted 
mass vanishes.
For $z'<0$, $\bar c < 1$, corresponding to the condensate being
``twisted'' by the twisted mass. Analytically continuing to
imaginary $\mu$ corresponds to staying on the same root for
positive $z'$. One finds that this root joins with another
and becomes two complex roots at $z'\approx 0.2$. This leads
to exactly the expected analytic structure,
namely a branch cut on
the positive real $z'$ axis, as one can verify by expanding
the function in the vicinity of the position where the roots join.
One can thus read off the gap: $\bar\alpha \approx 0.2 m_\pi^4/(2 B_0)^2$.
This is significantly smaller
than the result one would obtain ignoring discretization errors,
$\bar\alpha_{\rm cont} = m^2 = m_\pi^4/(2 B_0)^2$
(corresponding to $z'=1$).

\begin{figure}
\centering
\subfigure[]{
\label{fig:reroot10}
\scalebox{1}[1]{\includegraphics[width=3in]{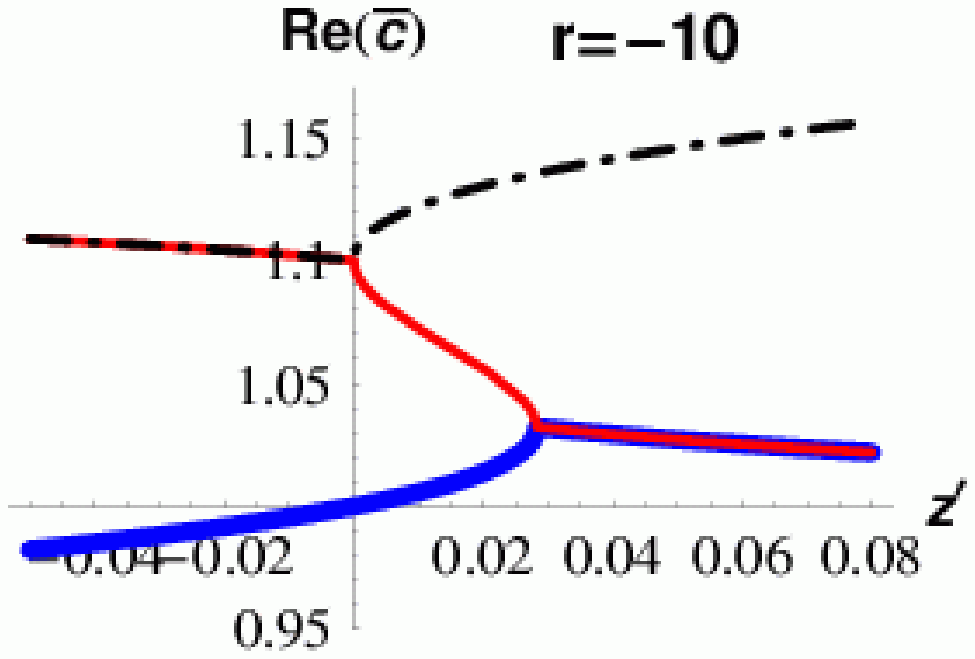}}}
\subfigure[]{
\label{fig:reroot.1}
\scalebox{1}[1]{\includegraphics[width=3in]{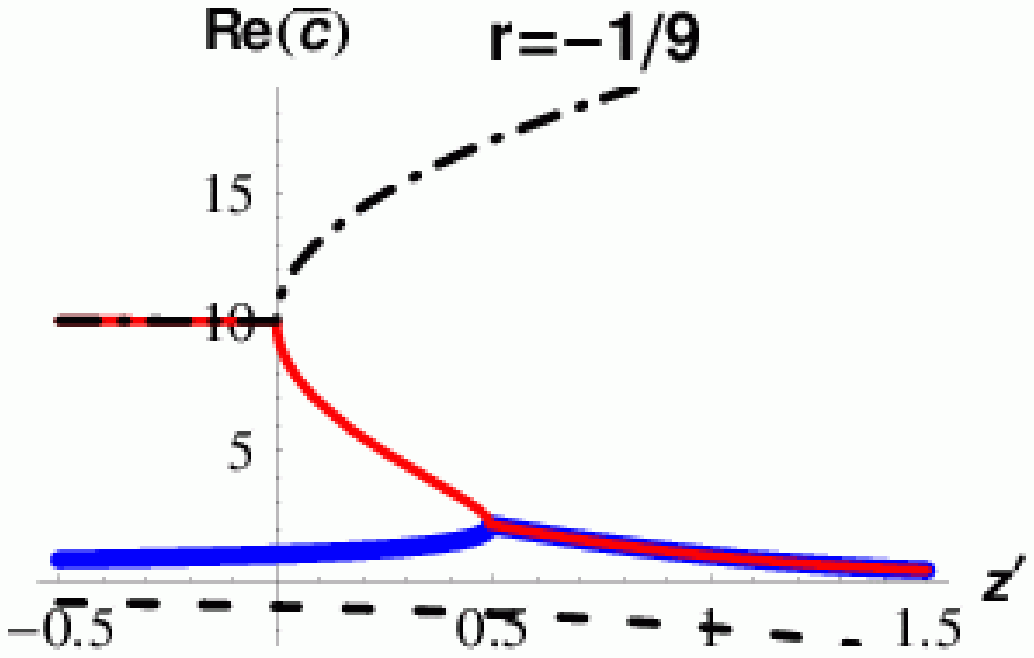}}}
\caption{\label{fig:reroot10.1}
\footnotesize
Real parts of roots for (a) $r=-10$ and (b) $r=-1/9$. Notation
is as in fig.~\protect\ref{fig:root1}. The vertical scale is chosen
to magnify the region of interest, so that not all roots appear in
(a).}
\end{figure}

\begin{figure}
\centering
\subfigure[]{
\label{fig:reroot.01}
\scalebox{1}[1]{\includegraphics[width=3in]{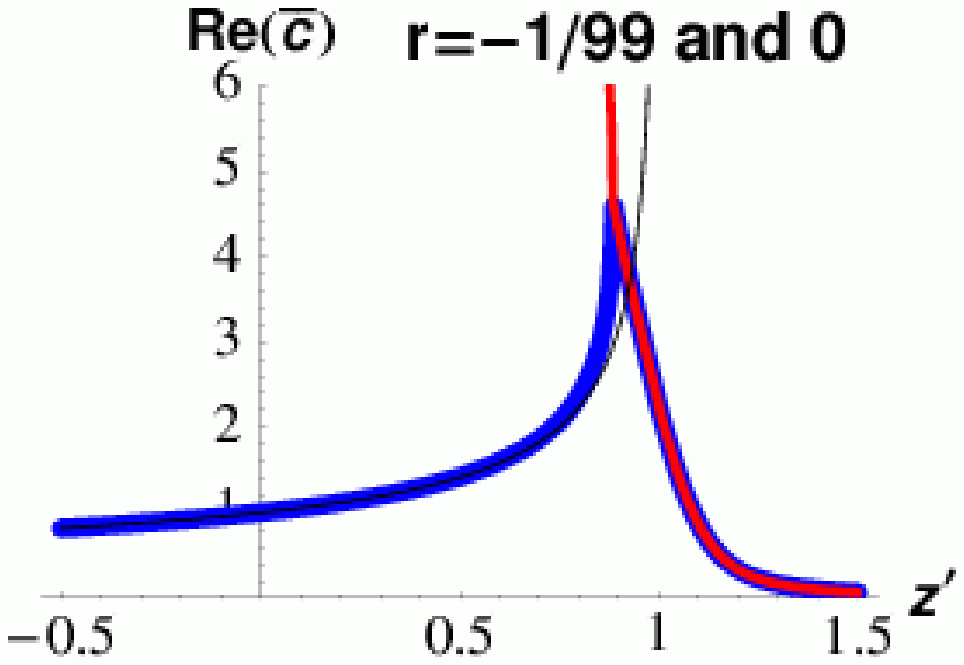}}}
\subfigure[]{
\label{fig:imroot.01}
\scalebox{1}[1]{\includegraphics[width=3in]{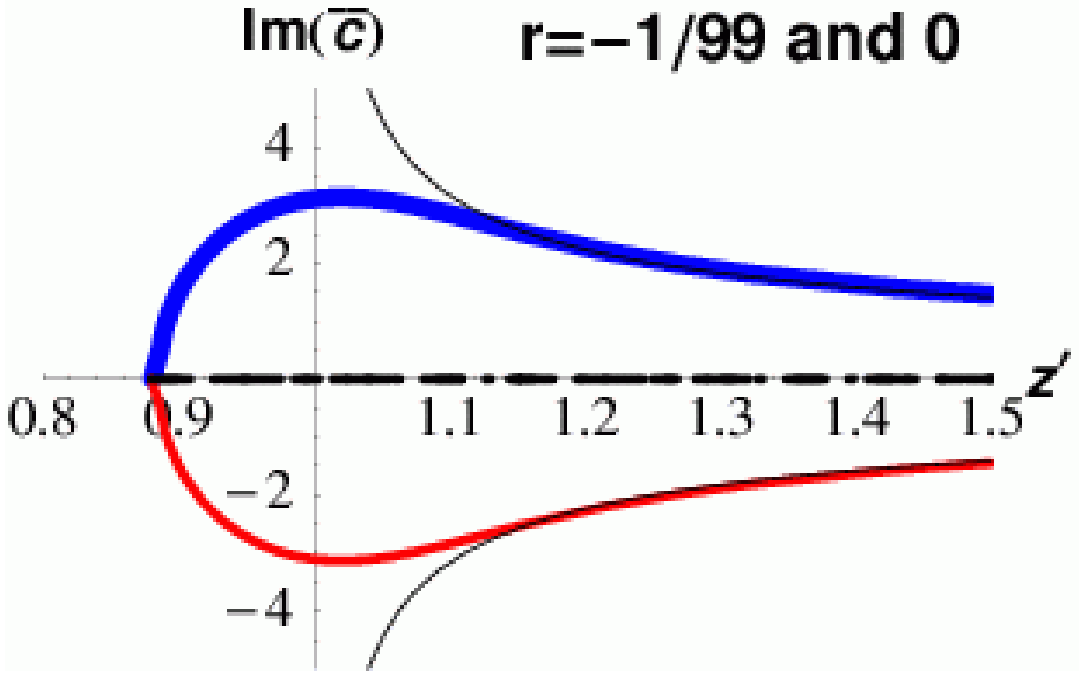}}}
\caption{\label{fig:root.01}
\footnotesize
Real and imaginary parts of roots for $r=-1/99$ 
(with notation as in fig.~\protect\ref{fig:root1}, except not
all roots appear given the vertical scales) compared to those in
the continuum limit, $r=0$ (shown as a thin solid [black] line). 
Note that the horizontal scales differ in the two plots. }
\end{figure}

As one moves away from the continuum limit, the discretization errors
in the gap become more pronounced. Figure~\ref{fig:reroot10} shows the
roots for $r=-10$, for which the gap is $\approx 2\%$ of the continuum
result. By contrast, fig.~\ref{fig:reroot.1} ($r=-1/9$) shows
the gap moving towards its continuum value as one reduces the discretization
errors. Very close to the continuum ($r=-1/99$),
the gap is almost unity, 
and the roots track those for $a=0$ except close to the
gap, as illustrated in fig.~\ref{fig:root.01}.

\begin{figure}
\centering
\subfigure[]{
\label{fig:gap}
\scalebox{1}[1]{\includegraphics[width=3in]{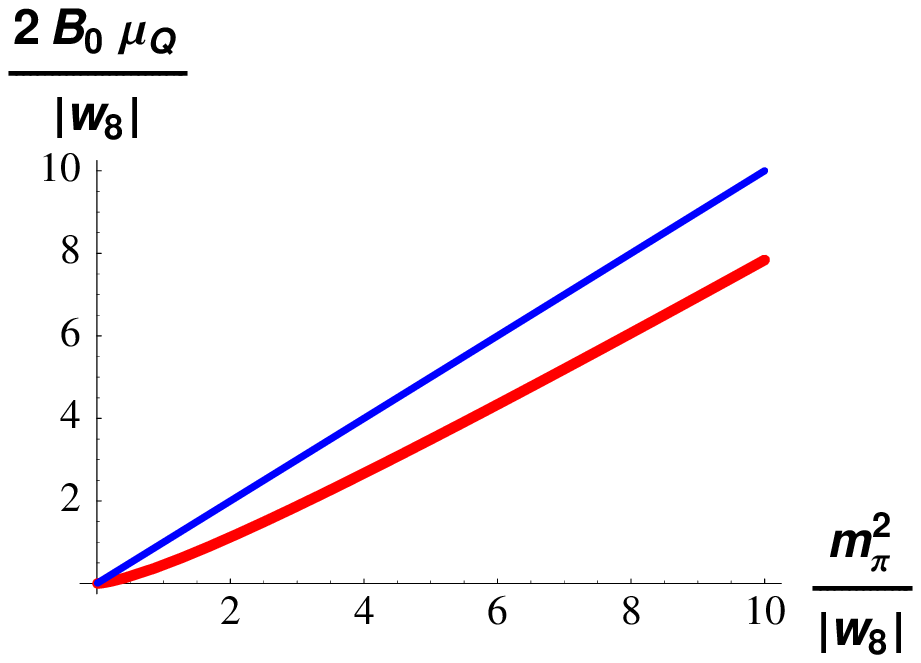}}}
\subfigure[]{
\label{fig:gaprat}
\scalebox{1}[1]{\includegraphics[width=3in]{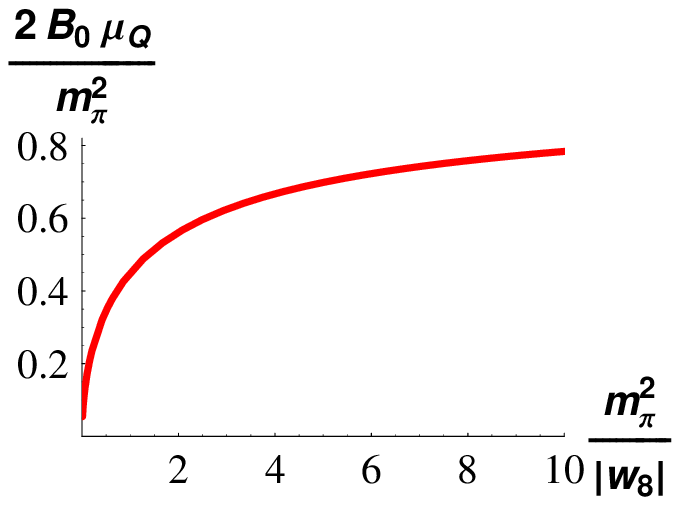}}}
\caption{\label{fig:gapandrat}
\footnotesize
Results for the spectral gap of $Q_m$, $\mu_Q$, plotted against
$m_\pi^2/|w_8|$.
(a) Gap in dimensionless units, $2 B_0 \mu_Q/|w_8|$ (thick [red] curve)
and result in continuum limit (thin [blue] curve);
(b) Ratio of gap to $m_\pi^2/(2 B_0)$, which is unity in continuum limit. 
Results are for $w_8<0$. }
\end{figure}

I next show how the gap depends on the pion mass-squared at fixed $a$.
In order to facilitate comparison with the results of Ref.~\cite{DD}, I plot
the gap for $Q_m$ (which is called $\mu$ in Ref.~\cite{DD}, but which
I call $\mu_Q$ here to avoid confusion with the twisted mass). More precisely,
I plot, in fig.~\ref{fig:gap},
$2B_0 \mu_Q/|w_8|=2B_0\sqrt{\bar\alpha/w_8^2}$
versus $m_\pi^2/|w_8|$. Dividing by $|w_8|$ gives 
dimensionless quantities.
If $|r_{68}|\ll 1$, the $x$-axis gives the pion mass-squared 
in units of the Aoki-phase half-length.
The quantities are such that, in the continuum limit, 
the result is a straight line with unit slope.

The predicted gap is smaller than the continuum expectation. To bring this out,
I show in fig.~\ref{fig:gaprat} the ratio of the gap to the continuum result,
$m_\pi^2/(2 B_0)$. This approaches unity for $m_\pi^2/|w_8| \gg 1$, but the
discretization errors reduce it significantly as one approaches the 
Aoki-phase end point.
The analytic form of the result as $m_\pi^2\to 0$ is
\begin{equation}
\frac{2 B_0 \mu_Q}{m_\pi^2} = \frac{2 B_0\sqrt{\bar\alpha}}{m_\pi^2}
= \left(\frac23\right)^{3/2} \sqrt{\frac{m_\pi^2}{|w_8|}}
\,.
\label{eq:endpoint}
\end{equation}
Note that the vanishing of the gap at the end point
{\em does not imply} that the ratio of fig.~\ref{fig:gaprat} must vanish too. 
It would be sufficient if the ratio tended to a finite constant as $m_\pi^2\to0$.
Thus discretization errors lead to an additional reduction in the gap, which
makes simulations less ``safe'' in the sense of Ref.~\cite{DD}. The 
square-root behavior is related to the mean-field exponents that apply at LO. 
In fact, if one approaches too close to the end point, higher order
terms in \chpt\ become important, and the exponents
are corrected~\cite{Aoki03,observations}.

\begin{figure}
\begin{center}
\epsfysize=4truein 
\epsfbox{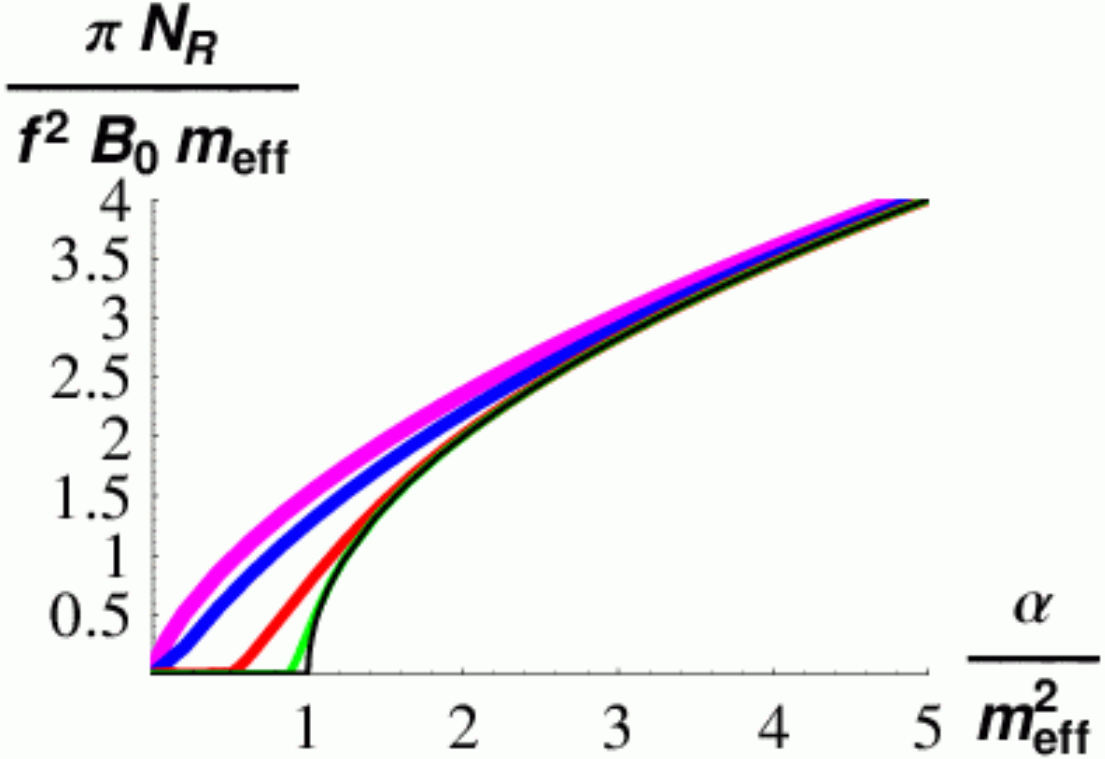}
\end{center}
\caption{\footnotesize
Results for integrated spectral density $N_R(\alpha)$. I plot
$[\pi/(B_0 f^2 m_\eff)] \times N_R(\alpha)$  
versus $\alpha/m_\eff^2$, where $m_\eff=m+w_6/B_0=\hat{m}^\eff/(2B_0)$.
Moving from top to bottom, results are for $r=-10$ (magenta), 
$-1$ (blue), $-1/9$ (red), $-1/99$ (green) and $r=0$ (black).
The latter is the continuum result if $w_6=0$.
Note that the upper two curves do have a gap (i.e. $N_R(\alpha)$
only becomes non-zero starting at a non-zero value of $\alpha$), 
but this is not visible in the figure.
}
\label{fig:spect}
\end{figure}

The method also yields the discretization errors in
the integrated spectral density. I show the impact of these errors for
the standard values of $r$ in fig.~\ref{fig:spect}. The 
units of $N_R(\alpha)$ are such that the continuum result
(\ref{eq:contNR}) is\footnote{%
The analytic form as $r\to-\infty$ asymptotes, for small
$\alpha/m_\eff^2$, to $(\pi N_R)/(B_0 f^2 m_\eff) = (3\sqrt{3}/2)
(v/2)^{2/3}$.}
\begin{equation}
\frac{\pi}{B_0 f^2 m_\eff} \times N_R(\alpha)
= 2 \sqrt{\alpha/m_\eff - 1}
\,.
\end{equation}
The figure shows what happens if $m$ is held fixed
and the lattice spacing reduced.
If $|w_6|\ll |w_8|$, then far above the gap
 the integrated density is unaffected by discretization errors,
while there are significant downward shifts at the low end of the spectrum.
In other words, aside from the smallest eigenvalues, say those
with $\alpha < 3 m^2$, the spectrum is
that of the continuum using the mass $m$. Note that this is {\em not} the
quark mass which vanishes at the Aoki-phase end point (where $m_\pi\to 0$).
It is the so-called PCAC mass,
which vanishes instead in the center of the Aoki phase.
The effect of a non-vanishing $w_6$ is essentially to shift the
spectrum as a whole by an amount of $O(a^2)$.

\bigskip
I now explain the failure of the method for $w_8>0$. 
It turns out that the nature of the failure depends
on the sign of $\hat{m}^\eff=\hat{m}+2w_6$. 
As can be seen from fig.~\ref{fig:w6w8B},
the boundary between positive and negative $\hat{m}^\eff$ is
also the line along which $r=1$.

\begin{figure}
\centering
\subfigure[]{
\label{fig:rerootwposa}
\scalebox{1}[1]{\includegraphics[width=3in]{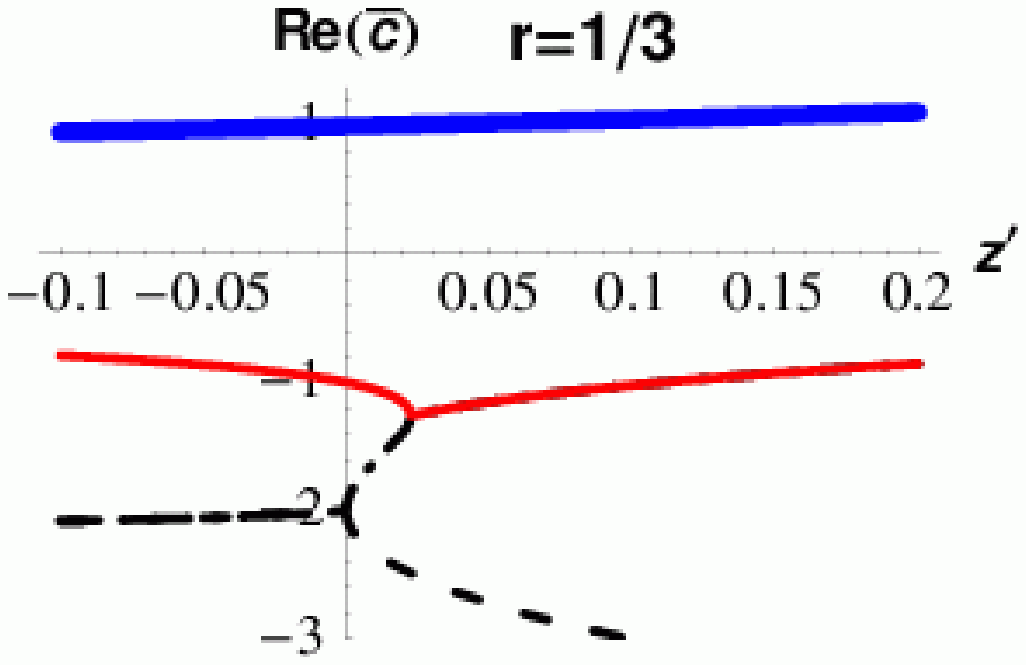}}}
\subfigure[]{
\label{fig:rerootwposb}
\scalebox{1}[1]{\includegraphics[width=3in]{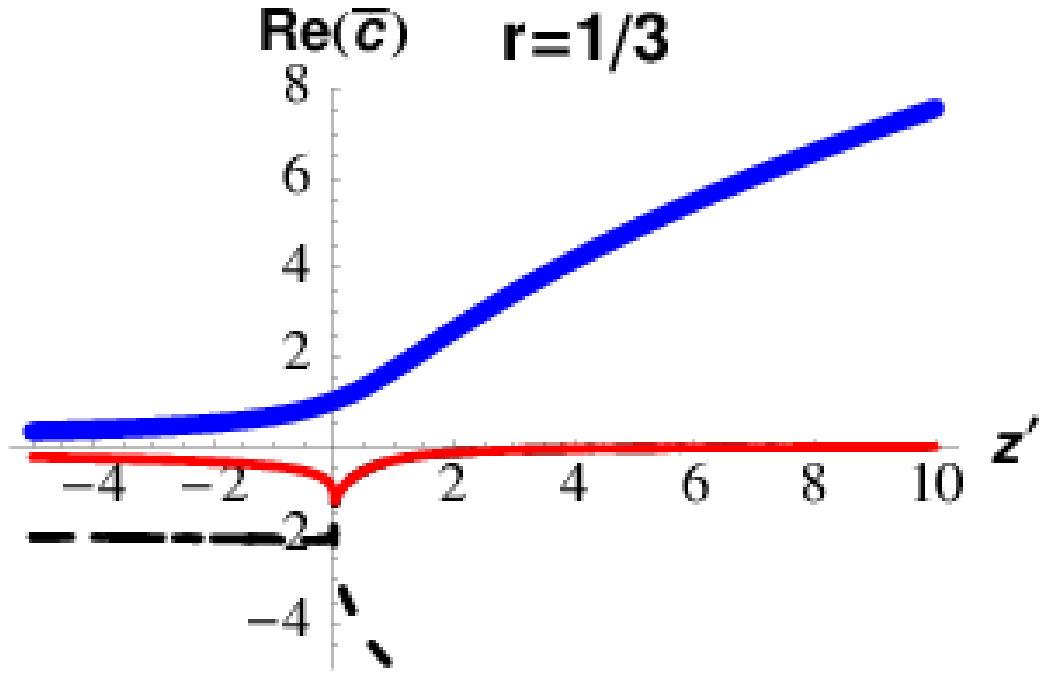}}}
\caption{\label{fig:rerootwpos}
\footnotesize
Real part of roots for $r=1/3$, for two choices
of scale of $z'$. The root of interest is shown
by the thick (blue) solid line. 
}
\end{figure}

I first consider $\hat{m}^\eff \ge 0$ which, given that $w_8>0$, 
implies that $0 < r \le 1$. For $|r_{68}|<1$ this
corresponds to the first-order scenario.
I show the real part of 
the roots for a representative choice, $r=1/3$,
in fig.~\ref{fig:rerootwpos}. This corresponds to
$r'=2$ if $r_{68}=0$, i.e. discretization errors and the
quark mass contribute about equally to the pion mass.
The key point to note is that the root of interest, which passes
through $\bar c=1$ at $z'=0$, is real for all real choices of $z'$.
This remains true throughout the range $0<r < 1$.

\begin{figure}
\begin{center}
\epsfysize=4truein 
\epsfbox{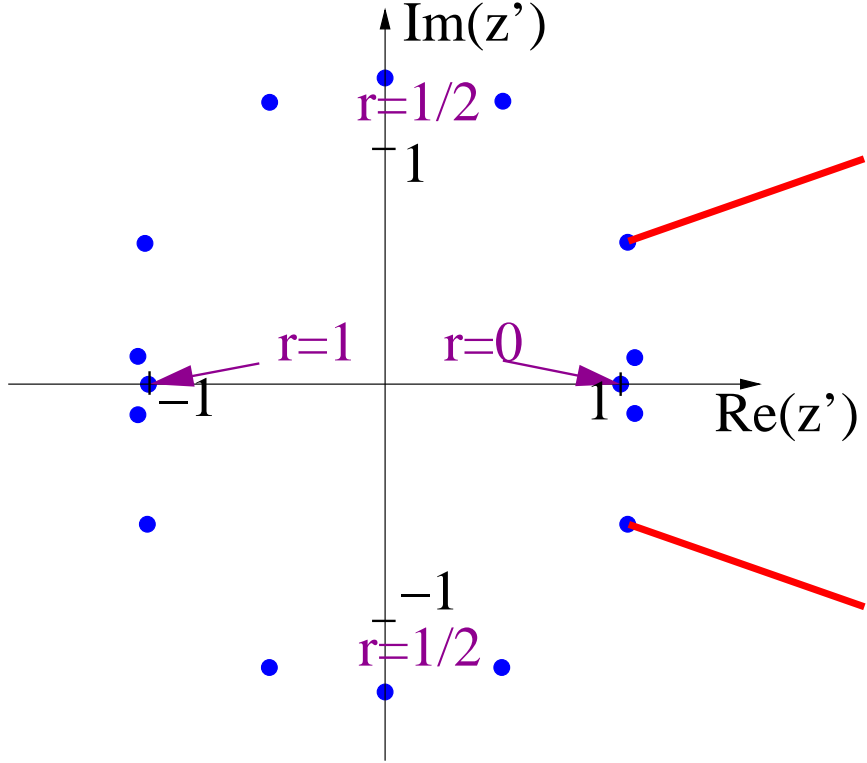}
\end{center}
\caption{\footnotesize
Analytic structure of $\CV^\aux_\mu(z')$
for $w_8>0$ and $\hat{m}^\eff > 0$, corresponding
to $0<r<1$. The cuts (thick [red] lines)
are shown for the example of $r=0.1$.
As usual, the position of the
cuts is a convention, while their end points
are fixed. The trajectory of the end points as
$r$ changes is shown by the (blue) dots. The values of
$r$ chosen are $0$, $0.01$, $0.1$, $1/3$, $1/2$, $2/3$, $0.9$,
$0.99$ and $1.0$. Note the symmetry under $r \to 1-r$,
$Re[z']\to - Re[z']$.
}
\label{fig:cut}
\end{figure}

This result implies that there is no discontinuity
in the potential, and, using eq.~(\ref{eq:final}), that 
the integrated spectral density vanishes at LO. This is a peculiar result,
particularly because, by sending $w_6,w_8\to 0$
(implying $r\to 0^+$)  one can approach
arbitrarily close to the continuum, where the method works and
one obtains a non-vanishing spectral density.
Mathematically, what happens is as follows. Although the
root of interest does not join with others for real $z'$, it
does so at a pair of complex conjugate values. Thus there
are two cuts in the complex plane, as illustrated in
fig.~\ref{fig:cut}. As $r\to 0^+$, the two cuts come together
at $z'=1$, leading to a single cut on the real axis,
given by the continuum result eq.~(\ref{eq:contF}). 
As $r$ varies between $0$ and $1$,
the ends of the cuts follow the paths in the complex plane sketched
in fig.~\ref{fig:cut}. They come together again when $r=1$, leading
to a cut along the negative $z'$ real axis starting at $z'=-1$.\footnote{%
The appearance of a cut on the negative real axis for $r=1$
can be understood if $w_6=0$.
Then $r\to1^-$ corresponds to $w_8\to +\infty$,
which means, since $\hat{m}$ is held fixed,
that one is approaching the first-order transition boundary.
But in this case there is a non-analyticity, starting at $\hat{\mu}=w_8=m_\pi^2$,
namely the end point of the first-order transition line.
I do not know of a physical explanation of the presence of 
this cut if $w_6\ne 0$.}

The implication of these results is that for $w_8>0$ and $\hat{m}^\eff\ge 0$
the analytic structure of the root
of interest, which is inherited by the potential $\CV^\aux_\min$,
differs from that of $F(z)$. Recall that the latter has,
by construction,  a cut only
along the positive real axis, starting at $\bar\alpha>0$.
Thus the result derived above relating $F$ to $\CV^\aux_\min$,
eq.~(\ref{eq:FtoCV}), fails.
The conclusion I draw is that the LO approximation 
is inconsistent for $F(z)$ if $w_8>0$, and higher order terms
are needed, a possibility already discussed following
eq.(\ref{eq:powercount}). An alternative is that the gap vanishes, so that
the primary method is inapplicable, but this is implausible
as it does not match onto the continuum result as $a\to0$.

\begin{figure}
\centering
\subfigure[]{
\label{fig:reroot2}
\scalebox{1}[1]{\includegraphics[width=3in]{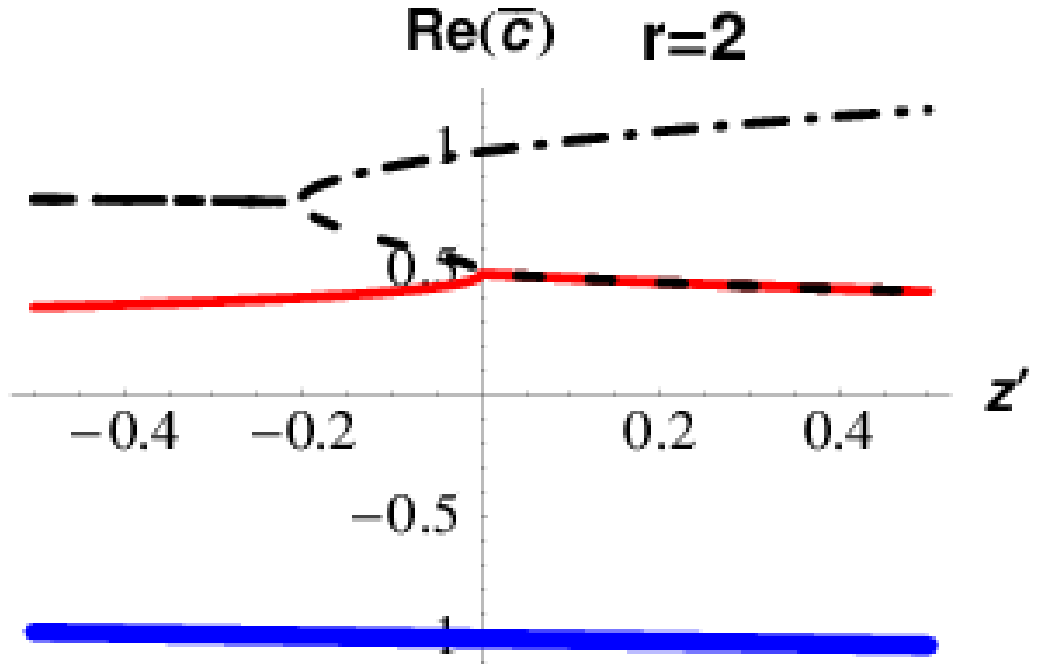}}}
\subfigure[]{
\label{fig:imroot2}
\scalebox{1}[1]{\includegraphics[width=3in]{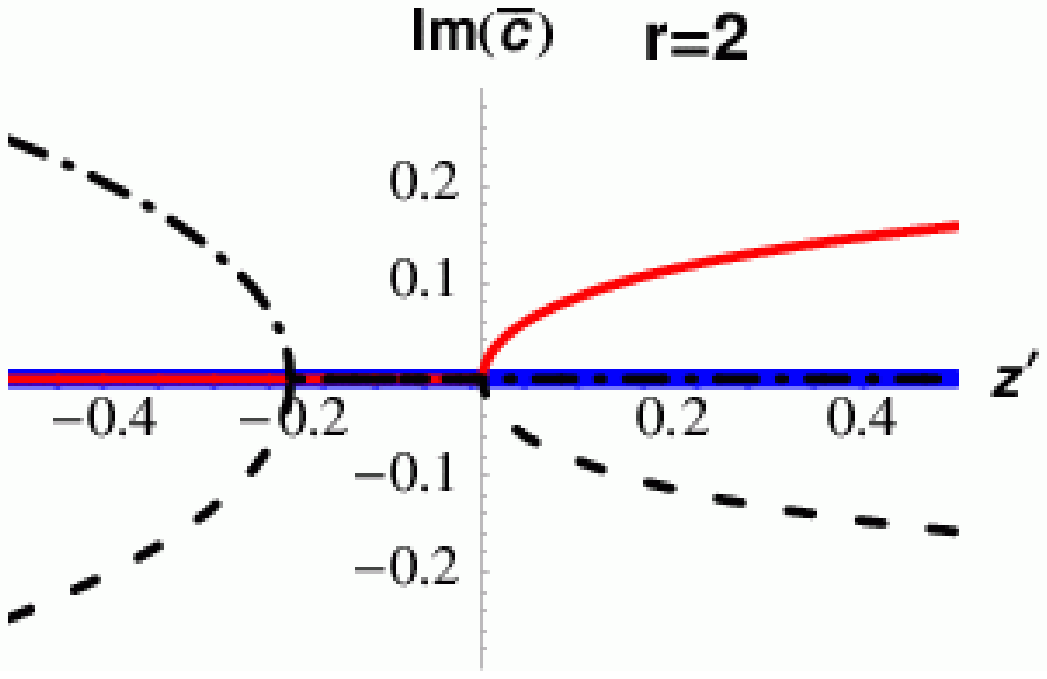}}}
\caption{\label{fig:root2}
\footnotesize
Real and imaginary parts of roots for $r=2$. The root of interest
is that shown by the thick (blue) line.} 
\end{figure}

The method fails for essentially the same reason when
$w_8>0$ and $\hat{m}^\eff < 0$. This corresponds 
to $1< r <\infty$, and, in fig.~\ref{fig:w6w8B}, is 
the region below the horizontal $r=1$ line. The generic behavior
of the roots is illustrated by the example of $r=2$, shown
in fig.~\ref{fig:root2}. Because $\hat{m}^\eff<0$ the
root which minimizes the potential for real $\mu$ is that for
which $\bar c=-1$ at $z'=0$. For $1<r<\infty$, this root
does not connect with any others for real $z'$. The situation
is very similar to that for $\hat{m}^\eff >0$, namely there are two
cuts in the complex plane. As before, since this analytic
structure differs from that of $F(z)$, it must be that
the approximations leading
to the relation (\ref{eq:FtoCV}) break down.

\bigskip
Despite the failure of the method for $w_8>0$, there is no reason
to doubt the validity of results obtained for $w_8<0$.
These results provide a self-consistent solution for $F(z)$ and
thus for the resolvent $R_R(z)$. What I mean by this is that the
series for $R_R(z)$, evaluated term by term at LO, and summed,
gives an analytic function with a cut {\em only} on the real
$z$ axis, and no other singularities. This follows from
the properties of the roots of the quartic, but, to be sure, I
have checked this result numerically by verifying the
dispersion relation eq.~(\ref{eq:resolvent}) for real $z<\bar\alpha$.
I stress that the self-consistency of the LO result
does not imply that there are no higher order
corrections to $\rho_R(\alpha)$ and $R_R(z)$.

\section{Discussion and Outlook}\label{sec:conc}

In this paper I have attempted to generalize the methods
developed in the continuum for calculating the spectral density
of $-i\Dslash$ to the operator of interest for the lattice theory, $Q_m$.
The generalization necessary for the inclusion of discretization errors
turned out, at least in the approaches I have followed, to be rather involved
and only partly successful at leading-order.

Nevertheless,
the success of the method for $w_8<0$ leads to a result which is
relevant for numerical simulations, namely the
reduction in the spectral gap of $Q_m$ compared to the continuum expectation.
Since this likely applies only for the scenario in which discretization errors
lead to an Aoki phase, it is probably not directly applicable to the
numerical results for the gap obtained in Ref.~\cite{DD}. These are
obtained with the Wilson gauge and fermion actions, for which one is
apparently in the first-order scenario.

The result may be relevant, however, for simulations that are
presently underway using improved Wilson fermions and/or improved
gauge actions, for which the phase structure has not been established.
With this in mind, I briefly discuss the way in which the discretization
errors in the gap might best be seen. Looking at fig.~\ref{fig:gapandrat},
one sees that, for $m_\pi^2/|w_8|\gtapprox 2$ a linear fit of $\mu_Q$ to
$m_\pi^2$ (or, at this order, equivalently to $m$) would work reasonably well,
but there would be a non-zero, negative intercept, and the slope would be
smaller than the continuum expectation. Significant curvature should appear
below this point. If one were in the Aoki-phase scenario with $w_6\approx 0$
and $|w_8|$ having a numerical value close to that which explains the
first-order transition observed with Wilson gauge and fermion actions,
i.e. if $w_8=-a^2\Lambda^4$ with $\Lambda\approx 700 MeV$, then this
``critical mass'' is $m_\pi=250$ and $150\;$MeV at $a=0.1$ and $0.06\;$fm,
respectively. Of course, if $\Lambda$ were reduced by a factor of 2, as might
be achieved by using alternate actions, these values for $m_\pi$ would be reduced
by a factor of four.

It is important to note that, since the discretization effects discussed here
are of $O(a^2)$, there is no {\em a priori} reason to expect them to be smaller
for improved quark actions. One must simply investigate the issue numerically.

On the theoretical side, 
the failure of the primary method at LO for $w_8>0$ is puzzling.
One can imagine working at arbitrarily small $a$ and $m$ where one would
expect the corrections to the LO form to become arbitrarily
small, and yet, apparently, they must be able to lead to a significant
change in the analytic structure of the resolvent $R_R(\alpha)$.
In an attempt to understand this puzzle I have repeated the calculation
using a second method, and, although some new features
appear, they do not appear to provide a solution for general values
of the low-energy constants (and in particular for $w_6\approx w_7\approx 0$).

Despite this puzzle, I stress again that the result for $w_8<0$ provides
a self-consistent solution for $R_R(\alpha)$, and I see no reason to doubt
this result.

Beyond the task of resolving the above-mentioned puzzle,
there are a number of directions in which it would be interesting to pursue
this work. It could be applied for non-zero twisted sea-quark
mass, although this
is perhaps more of theoretical than practical interest as the gap is 
known to be bounded by the twisted mass. Of more practical interest would
be to determine the impact of working at finite volume. I expect the
corrections to the results for the spectral density to be exponentially suppressed
until one enters the $\epsilon$-regime. As noted in the introduction, however,
fluctuations in the gap have been observed to fall as inverse powers of $V$,
and it would be interesting to see whether the effective theory 
can reproduce this behavior.

\section*{Acknowledgments}
I am very grateful to Oliver B\"ar, Maarten Golterman,
Martin L\"uscher, Ruth Van de Water, 
and my colleagues at the University of Washington
for comments and discussions.
This research was supported in part by 
U.S. Department of Energy Grant No. DE-FG02-96ER40956.

\appendix

\section{Derivation of key result}
\label{app:key}

In this appendix I demonstrate the result in eq.~(\ref{eq:key}),
which I repeat here for clarity:
\begin{equation}
2 (n-1)! \langle P_R^n \rangle_{PQ,LO}
= 
\langle(P_{R,12}+P_{R,21})^n\rangle_{CONN,LO}
\,,\qquad
(n\ge 1)
\,.
\label{eq:keyapp}
\end{equation}
As noted in the text, this result is trivial if $n$ is odd,
so I assume that $n$ is even in the following. Recall that
both correlators are to be calculated using the auxiliary
chiral potential $\CV^\aux_\chi$ of eq.~(\ref{eq:Leff'}),
in which there are only single-supertrace operators.

The demonstration is based on the following results.
\begin{enumerate}
\item[(I)] {\em Only charged pion propagators appear in 
$\langle(P_{R,12}+P_{R,21})^n\rangle_{CONN,LO}$
and $\langle P_R^n\rangle_{PQ,LO}$.}

Charged here means being flavor off-diagonal,
e.g. ``12'' or ``35'', while
neutral means flavor diagonal: ``11'', ``22'' \dots .
The argument goes as follows. Tree-level diagrams can always be
divided into two pieces by cutting any propagator.
Assume that there is a neutral pion propagator and cut it.
Since the operators $P_R$ are all charged, there must be an even
number in each of the two resulting pieces to add up to an overall
neutral combination.
But an even number of $P_R$'s
can, together, produce only an even number of pions, which, through
interactions, can only change to an even number of pions.
Thus one reaches a contradiction. 

In fact, for
$\langle P_R^n\rangle_{PQ,LO}$ the argument is more simple:
the only set of pseudoscalar operators which are neutral is the
whole set, but that leaves no operators for the other piece
of the diagram.

\item[(II)]
{\em For each pionic diagram contributing to the correlators
of interest, one can unambiguously associate a finite number of
planar quark-flow diagrams.}

This holds true as long as one does not
have neutral pion propagators,
as is the case here given observation (I).\footnote{%
The potential problem with neutral propagators is that
they can have ``hidden'' quark-disconnected contributions due to
the constraint $\Str(\Phi)=0$. Thus one does not know
how to assign the quark-flow without further considerations.
In fact, in the present case, with all valence quarks degenerate, 
this restriction can be shown to be unnecessary,
but it simplifies the argument.}
To obtain the quark-flow diagram
one simply follows the flavor indices, connecting them
into a single ``trace'' at each vertex and at each pseudoscalar
operator.\footnote{%
Although it is not needed for the argument, I note that
quark-flow can also be traced if there are vertices
with multiple supertraces.}
The flows are directed, with arrows indicating the flow of
fermion number. Note that I do not distinguish between diagrams
in which the flavor labels are permuted: these are all lumped into
a single quark-flow ``skeleton'', with the counting of possible
attachments of flavor labels to be done below.

For each pion-level diagram
there are, in general, multiple quark-flow diagrams,
corresponding to different orderings of the
contractions between the $P_R$.
For example, the pion-level diagram of fig.~\ref{fig:6pt}c
leads to the three quark-flow diagrams of
fig.~\ref{fig:4ptqline}b-d.
I have drawn the diagrams in fig.~\ref{fig:4ptqline} to be planar,
in the sense that no quark-lines cross. This is their canonical
form, for then the quark-line runs around the outside of the
diagram, and the connection of flavor indices is transparent.
That it is always possible to draw the quark-flow diagrams in
this way is property of tree-level diagrams, following
from the fact, noted above, that tree diagrams fall into
two pieces if one cuts any propagator. Thus one can cut all
propagators, make each vertex and insertion of $P_R$ planar,
and then glue the propagators back together 
starting from an arbitrary point in the diagram, resulting in 
an overall planar diagram.
The key point is that the vertices and insertions can be treated independently.

\item[(III)]
{\em All contributions to $\langle P_R^n\rangle_{PQ,LO}$
contain a single quark-line in the corresponding quark-flow diagrams.}

For each quark-line in the diagram, all of which are necessarily closed,
the total ``charge'' of the $P_R$ attached to it must vanish.
Given the charges in $\langle P_R^n\rangle_{PQ}$ (see eq.~\ref{eq:MkPQ}),
this is only possible if all the $P_R$ are attached to a quark-line,
so there can only be one such line.
Examples are shown in fig.~\ref{fig:4ptqline}.
This result implies the (rather obvious) fact that there are
no disconnected contributions to $\langle P_R^n\rangle_{PQ,LO}$.

\item[(IV)]
{\em All contributions to
$\langle(P_{R,12}+P_{R,21})^n\rangle_{CONN,LO}$
contain a single quark-line in the corresponding quark-flow diagrams.}

This result relies on the facts that $\CV^\aux_{\chi}$ contains only
single-supertrace operators and that there are no neutral propagators
[observation (I)].
These facts imply that there is no way for a quark-line to ``close off'' within
a vertex or a propagator. (Examples of how this is possible with
two-supertrace operators are given in fig.~\ref{fig:twostrace}a-b.)
Thus, if there is more than one quark-line, the pion diagram must fall
into as many pion-disconnected pieces as there are quark-lines.
This cannot happen, by definition, if the diagram is 
connected at the pion-level.

Note that this result fails beyond LO. A one-loop example of a contribution
to $\langle(P_{R,12}+P_{R,21})^4\rangle_{CONN}$ which involves two quark-lines,
and is thus absent in $\langle P_R^4\rangle_{PQ}$, is given in
fig.~\ref{fig:oneloop}.

\begin{figure}
\begin{center}
\epsfysize=1truein 
\epsfbox{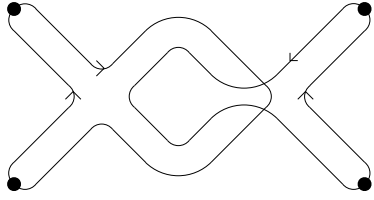}
\end{center}
\caption{\footnotesize
Example of quark-flow for a one-loop
diagram involving two single-supertrace operators
which cannot contribute to $\langle P_R^4\rangle_{PQ}$ but which does
contribute to $\langle(P_{R,12}+P_{R,21})^4\rangle_{CONN}$.}
\label{fig:oneloop}
\end{figure}

\item[(V)] {\em The same planar quark-line diagrams contribute to
$\langle(P_{R,12}+P_{R,21})^n\rangle_{CONN,LO}$
and $\langle P_R^n\rangle_{PQ,LO}$.}

Any pion-level diagram contributing to 
$\langle(P_{R,12}+P_{R,21})^n\rangle_{CONN,LO}$
can also contribute to $\langle P_R^n\rangle_{PQ,LO}$,
and vice-versa,
since the vertices and insertions of $P_R$ produce the same
number of pions in both cases. Thus the pion-level diagrams are
the same for both correlators.

The remaining question is whether 
the quark-flow diagrams that contribute to a given
pion-level diagram are the same for both correlators.
That this is the case can be seen as follows. From 
(III) and (IV),
all quark-line diagrams contributing to both correlators
involve a single quark-line. Adding back in the flavor indices,
these will alternate as $1212\dots2$ in the unquenched correlator,
and as $1234\dots(2n)$ in the PQ correlator, as illustrated in
fig.~\ref{fig:UQvsPQ}.
Simply by interchanging these flavor labels 
one interchanges legitimate quark-line diagrams in
the two correlators. Thus if a quark-flow diagram is present
in one correlator, it will be present in the other.

\item[(VI)] {\em The relative contribution from each
planar quark-flow diagram to $\langle(P_{R,12}+P_{R,21})^n\rangle_{CONN,LO}$
and $\langle P_R^n\rangle_{PQ,LO}$ is the relative
contraction factor $2(n-1)!$.}

The contributions to both correlators have now been broken
down to those from a sum over (the same) planar quark-line diagrams.
These diagrams are nothing more than an explicit representation of
all the ways the flavor indices are connected when the pion fields are
contracted. Thus all flavor factors are accounted for explicitly.
The other factors that contribute to the diagram are the propagators,
the factors which result from expanding out $\Sigma$ and $\Sigma^\dagger$
to obtain the vertices and insertions, and the counting of contractions.
Only the latter differ between PQ and unquenched correlators.
The propagators are the same because all quark masses are the same,
and the expansion of $\Sigma^{(\dagger)}$ leads to the same factors in both
theories. Furthermore, the ``internal'' counting of contractions---how many
ways the pion fields in the vertices can be contracted with each other to
yield the given planar quark-flow diagram---are also common.
The only part of the calculation which differs
between the theories is the choice of contractions with the external $P_R$. 
In the unquenched correlator,
there are $2\times  n!$ ways of associating the $n$ operators
$(P_{R,12}+P_{R,21})$ with the external insertions: $n!$ since each
operator can be attached to each insertion point, and $2$ because,
for each such attachment, one can globally interchange $P_{R,12}$ and
$P_{R,21}$. By contrast there are only $n$
contractions in the PQ case, because cyclicity must be preserved.
Thus the relative factor is $2 (n-1)!$, as claimed.

\end{enumerate}

In summary, by breaking the diagrams up into planar quark-flow parts,
which are common to both PQ and unquenched correlators, the problem
is reduced to counting contractions. The counting is the same for
all planar quark-flow diagrams with a given value of $n$, and thus
eq.~(\ref{eq:keyapp}) follows.


\begin{figure}[tb]
\begin{center}
\epsfysize=1.5truein 
\epsfbox{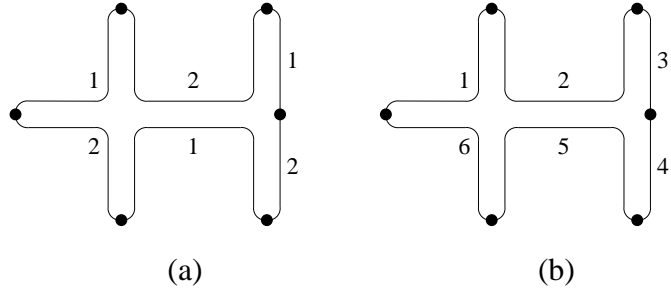}
\end{center}
\caption{\footnotesize
Examples of flavor labels for
a planar quark-flow diagram contributing
to fig.~\protect\ref{fig:6pt}c for (a) 
$\langle(P_{R,12}+P_{R,21})^6\rangle_{CONN}$, and
(b) $\langle P_R^6\rangle_{PQ}$. 
}
\label{fig:UQvsPQ}
\end{figure}

\section{Calculation using alternate method}
\label{app:replica}

The aim of this appendix is to check the results, 
and to understand the limitations,
of the primary method, by repeating the calculation 
using an alternate method,
in which partial quenching is implemented using the replica trick.

The starting point is eq.~(\ref{eq:PQapproach}), which shows that one
can obtain the average of the positive and negative
spectral densities by determining the discontinuity in
the twisted valence condensate. To determine this condensate 
at LO using the replica trick
one needs to minimize the potential of a theory with
$N_S=2$ sea-quarks and $2 N_V$ replica quarks.
Note that this theory is unquenched, and so has a
well-defined and bounded Hamiltonian. The use of $2 N_V$ replica
quarks, rather than $N_V$, is for later convenience.
Having calculated the condensate at the minimum, 
one then sends $N_V\to 0$, giving the valence condensate,
and finally analytically continues to complex $\mu$ to determine
the discontinuity across the imaginary axis.

The LO chiral Lagrangian takes the same form as that in the PQ theory,
eq.~(\ref{eq:LPQ}), except that $\Sigma\in SU(N_S+2N_V)$ and supertraces
are replaced by traces.
The potential is thus
\begin{equation}
\frac{\CV_\chi^\rep(\mu)}{f^2}
=
- \frac{\Tr\left(\chi^\rep_+\Sigma^\dagger+\Sigma\chi^\rep_-\right)}{4}
- w_6 \frac{\left[\Tr\left(\Sigma+\Sigma^\dagger\right)\right]^2}{16}
- w_7  \frac{\left[\Tr\left(\Sigma-\Sigma^\dagger\right)\right]^2}{16}
- w_8 \frac{\Tr\left(\Sigma^2 + {\Sigma^\dagger}^2\right)}{16}
\,.
\end{equation}
The mass term takes the block form
\begin{equation}
\chi^\rep_\pm={\rm diag}(\underbrace{\hat{m}}_{2\times 2{\rm\ sea\ block}}, 
\underbrace{\hat{m} \pm i\hat{\mu} \tau_3,\ \dots\ ,
\hat{m} \pm i\hat{\mu} \tau_3}_{N_V\ 2\times 2 {\rm \ blocks}})
\,.
\end{equation}
The twisted mass appears only
in the replica parts of the mass matrix since we are interested
in the spectral density of $Q_m$ for a sea sector with an 
untwisted mass.\footnote{%
One can now see how the continuum result (\ref{eq:LOcont}) in the main text
 is obtained.
Only the first term in $\CV_\chi^\rep$ is present, and $\chi_\pm^\rep
={\rm diag}(\hat{m}_S, \hat{m}_V, \dots, \hat{m}_V)$. Minimizing
the potential leads to the valence block of the condensate aligning
in the direction of $m_V$, independent of the alignment in the sea-quark
block.}
I recall that when using the replica trick one takes the LECs 
($w_6$ etc.) to be independent of $N_V$.

I assume that the form of $\Sigma$ which minimizes the potential
follows that of the mass matrix, i.e.
\begin{equation}
\Sigma={\rm diag}(\underbrace{1}_{2\times 2{\rm\ sea\ block}}, 
\underbrace{e^{i\omega \tau_3},\ \dots\ , e^{i\omega\tau_3}
}_{N_V\ 2\times 2 {\rm \ blocks}})
\,.\label{eq:replicacond}
\end{equation}
This can be shown to lead to a local minimum under certain
assumptions on the parameters, as I discuss at the
end of the appendix.
With the condensate given by (\ref{eq:replicacond}) 
the potential becomes
\begin{eqnarray}
\frac{\CV_\chi^\rep(\mu)}{f^2}
&\!=\!& 
- \hat{m} (1\! +\! N_V \cos\omega) - \hat\mu N_V \sin\omega
- w_6 (1 \!+\! N_V \cos\omega)^2 - \frac{w_8}{4} (1\! +\! N_V \cos 2\omega)\\
&\!=\!&
{\rm constant}- N_V\left[ \hat{m}^\eff \cos\omega + \hat\mu\sin\omega
+\frac{w_8}{4}\cos2\omega\right]
- N_V^2 w_6 \cos^2\omega
\,.\label{eq:replicapot}
\end{eqnarray}
Note that the $w_7$ term does not contribute, and that the same combination
$\hat{m}^\eff = \hat{m}+ 2 w_6$ appears as in the main text.
As $N_V\to 0$, the term quadratic in $N_V$ can be dropped, and
the determination of the minimum is to be done using only the
contribution linear in $N_V$. This contribution is exactly the 
auxiliary potential encountered in the primary method, eq.~(\ref{eq:Vauxvsomega}).
Thus the required minimizations in the two methods are identical,
and yield the same minimizing angle, $\bar\omega(\mu)$.

It remains to be shown that the results for the spectral density are
also the same. In the replica method, the required condensate is,
using eq.~(\ref{eq:Pchi}),
\begin{equation}
\langle \bar q_V \gamma_5 \tau_3 q_V\rangle_{LO}
= 2i B_0 f^2 \sin\bar\omega
\,.
\end{equation}
Since this is the condensate for {\em one} replica doublet
(and not $N_V$ doublets), it has a non-vanishing
limit as $N_V\to0$.
Thus the final result from the replica trick is, using
eq.~(\ref{eq:PQapproach}),
\begin{equation}
{B_0 f^2}\ {\rm Disc}\ [\sin\bar\omega]\bigg|_{\mu=i\lambda}
 = \pi \left[\rho_{Q}(\lambda)+\rho_{Q}(-\lambda)\right]
\,.
\label{eq:replicafinal}
\end{equation}
Note that the spectral density $\rho_Q$ is renormalized, since
it is determined from a renormalized condensate.

This result
is equivalent to that from the primary method, eq.~(\ref{eq:final}). 
To see this, one takes the derivative of the latter equation
with respect to $\alpha$, and uses the fact that the derivative
commutes with taking the discontinuity:
\begin{equation}
-2 i \pi \rho_R(\alpha) =
\frac{d}{d\alpha}\left\{
{\rm Disc}\;[\CV^\aux_\min(\mu^2=-z)]\bigg|_{z=\alpha}\right\}
= 
-{\rm Disc}\;[\frac{d}{d\mu^2}\CV^\aux_\min(\mu^2=-z)]\bigg|_{z=\alpha}
\,.
\label{eq:dfinal}
\end{equation}
Next, one notes that the spectral density of $Q_m^2$ 
can be related to that of $Q_m$,
\begin{equation}
{2 \sqrt{\alpha}}\; \rho_R(\alpha)
=
{\rho_{Q}(\sqrt{\alpha})+\rho_{Q}(-\sqrt{\alpha})}
\,,
\end{equation}
as well as the fact that
\begin{equation}
2 \mu \frac{d \CV_{\rm min}^\aux}{d (\mu^2)}
=
\frac{d \CV_\min^\aux}{d \mu}
=
- 2 B_0 f^2 \sin\bar\omega(\mu) 
\,.
\label{eq:relnIII}
\end{equation}
Substituting these results into eq.~(\ref{eq:dfinal}) and
performing some simple manipulations gives
eq.~(\ref{eq:replicafinal}).
The derivation also runs in the other direction: the
result in the text can be obtained from (\ref{eq:replicafinal})
as long as the discontinuity in $\sin\bar\omega$ is integrable,
which is the case here.

I now return to the question of whether the assumed form of
the condensate, eq.~(\ref{eq:replicacond}), is correct.
This brings up the general question of how to apply the replica
trick in a context where the dependence on $N_V$ is not perturbative.
It seems possible, for example, that, for a given choice of
LECs $w_{6-8}$, there could be a non-trivial $N_V$ dependence of
the phase structure, and thus a non-analytic dependence of
the condensate on $N_V$. This issue deserves further study,
which I have not attempted here. The only general comment I can make
is that, if $w_6=w_7=0$, then, for $w_8>0$ (which is the case in
which the primary method fails), it is plausible that
the condensate takes the form (\ref{eq:replicacond}) for all
integer $N_V$ and $\mu$ real.
This is because the mass term wants to align $\Sigma\propto \chi^\rep_+$,
while the $w_8$ term is minimized when $\Sigma=\pm 1$. These
competing ``pulls'' seem likely to lead to a condensate which lies
on the shortest path between the two favored directions,
which is eq.~(\ref{eq:replicacond}). Thus I do not expect that
studying the phase structure will  resolve the failures
of the primary method in general.

A more modest goal is to show that the form (\ref{eq:replicacond}) leads
to a local minimum. It is straightforward to show that 
minimizing the restricted replica potential (\ref{eq:replicapot})
does lead to an extremum in all directions of possible fluctuations.
Thus the issue is whether it is a minimum in all directions.
Choosing a basis in which there is no mixing, I find, in the
replica limit, that the pion mass-squareds are:
\begin{eqnarray}
M_{SS}^2 &=& \hat{m}^\eff+w_8
\,, \\
M_{VV,n}^2 &=& \bar c 
\left(\hat{m}^\eff + \frac{\hat{\mu}^2}{\hat{m}^\eff + w_8 \bar{c}}\right)
+{w_8}\; (2\bar{c}^2-1)
\,, \\
M_{VV,c}^2 &=& M_{VV,n}^2 + w_8\;(1 - \bar{c}^2)
\,, \\
M_{VS,n}^2 &=& M_{VV,n}^2 + w_7\; (1 - \bar c)^2
\,,\label{eq:MVSn}\\
M_{VS,c}^2 &=& \frac{M_{SS}^2 + M_{VV,n}^2 + w_8\; \bar c (1-\bar c)}{2}
\,.
\end{eqnarray}
Here $\bar c = \cos\bar\omega$, the subscripts 
``c'' and ``n'' stand for charged and neutral, respectively,
and the other subscripts indicate the sea and valence content of
the ``pions''.
If any of these mass-squareds passes through zero and becomes negative,
then the extremum is unstable to fluctuations in the corresponding
directions. Note that the alternate method has more directions of fluctuation
than appear in the primary method. Most notable
are the ``VS" directions---these introduce the LEC $w_7$ which does not
enter in the analysis of the main text.

If the alternate method is to resolve the puzzle from the main text,
it should do so for all choices of $w_7$, and so I begin with the simplest,
$w_7=0$. In this case, however, the results of the primary method are reproduced
(as long as $\hat{m}^\eff > 0$, an exception discussed below).
In particular,
if $w_8>0$ (and $\hat{m}^\eff>0$),  all pion mass-squareds are
positive for all real (positive and negative) values of $\mu^2$,
so that the choice (\ref{eq:replicacond}) with $\omega=\bar\omega$ is a local minimum.
If $w_8<0$, however, then $M_{VV,n}^2$ vanishes when $\mu^2=-\bar\alpha$.
This signals a non-analyticity in $\bar c$, which is inherited by the valence condensate,
and is identical to the condition for the gap discussed in the main text. 
Thus the puzzle is not resolved for these parameter choices.

There are, however, two cases where the present analysis differs from that in
the main text, both involving instabilities in which the mixed
pions composed of valence and sea-quarks have vanishing masses. 
The most straightforward follows from eq.~(\ref{eq:MVSn}):
as $\mu^2$ becomes more negative
$M_{VS,n}^2$ can pass through zero and become negative
if $w_7$ is sufficiently negative. 
It turns out that $w_7$ should not be too negative,
however, because then $M_{VS,n}^2$ also changes sign as $\mu^2$ becomes
more positive. Nevertheless, the conclusion is that there is a range
of negative $w_7$ which could lead to the expected analytic structure.

The other possibility is more complicated.
It occurs when $\hat{m}^\eff < 0$, i.e when $w_6$
is sufficiently negative ($2 w_6 < -\hat{m}$) that the mass
in the effective potential is negative even though $\hat{m}$ itself
is positive. Note that in this region 
one must have $w_8> |\hat{m}^\eff|$, as shown in fig.~\ref{fig:w6w8B}.
As noted in the main text,
if $\hat{m}^\eff < 0$,
the minimum of auxiliary potential
(or equivalently of $\CV_\chi^\rep$ for $N_V\to0$) 
lies at $\bar c=-1$ if $\mu=0$. 
The mass formulae given above
still hold, but now the contribution to $M_{VS,c}^2$ 
proportional to $w_8 \bar c(1-\bar c)$ is negative.
This has the effect that $M_{VS,c}^2$ vanishes exactly at $\mu^2=0$,
and becomes negative for $\mu^2<0$,
for all allowed $w_8$.
The vanishing at $\mu^2=0$ can be understood from the viewpoint
of the original replica potential, $\CV_\chi^\rep$. If $\bar c=-1$,
then the condensate takes the form 
\begin{equation}
\langle\Sigma\rangle = {\rm diag}(1, -1,\dots, -1)
\,.
\end{equation}
This breaks the $SU(2+2N_V)$ vector symmetry of the theory
that is exact when $\mu=0$. Thus, as in the Aoki-phase analysis,
there are exact Goldstone bosons, here the $VS$ pions.
This symmetry breaking arises because, for $w_6$ negative,
the $w_6$ term favors a condensate rotated away from $\pm 1$.

This instability seems to imply a discontinuity
in $\bar c$, and thus in the valence condensate, beginning at $-\mu^2=0$.
This would correspond to the gap vanishing.
If correct, this could explain the failure of the primary method,
which, unlike the present approach, relies on the gap being non-zero.\footnote{%
This illustrates a loophole in the claim made earlier in this Appendix
that the two methods are equivalent. 
This claim fails if the condensate rotates in directions other
than $VV$ or $SS$, because eq.~(\ref{eq:relnIII}) no longer holds.}
Note that one cannot reach the continuum limit in the region
of parameters where this possible resolution applies ($\hat{m}^\eff<0$,
$w_8> |\hat{m}^\eff|$), so there is no contradiction with the
result that the gap is non-vanishing in the continuum.

I have not pursued this analysis further, since the possible
resolutions of the problem for $w_8>0$ are not general.
In particular, they do not apply if 
$\hat{m}\gtapprox w_8 \gtapprox |2 w_6|$,
which is the region through which one likely
approaches the continuum limit if one is in the first order scenario.

\end{document}